%
%
\magnification=\magstep1
\baselineskip=11pt plus .1pt minus .1pt
\hsize=12.5truecm
\vsize=19.0truecm  
\hfuzz=5pt\vfuzz=5pt
\tolerance=1000
\overfullrule=0pt
\parskip=0pt
\abovedisplayskip=3 mm plus6pt minus 4pt
\belowdisplayskip=3 mm plus6pt minus 4pt
\abovedisplayshortskip=0mm plus6pt minus 2pt
\belowdisplayshortskip=2 mm plus4pt minus 4pt
\predisplaypenalty=0
\clubpenalty=10000
\widowpenalty=10000
\parindent=2em
%
%
\font\pgnumfont=cmr9
\font\headlinefont=cmti9
 \font\titlefont=cmbx10
\font\authorfont=cmr10
\font\addressfont=cmti9
\font\datefont=cmr9
\font\sumfont=cmr9
\font\itl=cmti9

\font\absfont=cmbx9
\font\secfont=cmr10
\font\subsecfont=cmti10
\font\subsubsecfont=cmr10
\font\figfont=cmr9
\font\figheadfont=cmbx9
\font\tabfont=cmr9
\font\tabheadfont=cmbx9
\font\mainfont=cmr10
\font\petitrm=cmr9

%
%
%
\newtoks\TITLE \newtoks\AUTHOR \newtoks\ADDRESS \newtoks\SUMMARY
\newdimen\sumindent \sumindent=\parindent
\newtoks\KEYWORDS \newtoks\SUBMITTED \newtoks\ACCEPTED
\newtoks\SENDOFF
%

%
%
\newtoks\firstpage
\let\firstpage=Y
\newtoks\AUTHORHEAD \newtoks\ARTHEAD \newtoks\VOLUME \newtoks\PAGES
\if!\the\AUTHORHEAD!\AUTHORHEAD={\the\AUTHOR}\fi
\if!\the\ARTHEAD!\ARTHEAD={\the\TITLE}\fi
\footline={\hfil}
\headline={\ifodd\pageno\rightheadline \else\leftheadline\fi}
\def\leftheadline{\if Y\firstpage\firsthead\global\let\firstpage=N
  \else\lefthead\fi}
\def\rightheadline{\if Y\firstpage\firsthead\global\let\firstpage=N
  \else\righthead\fi}
\def\lefthead{\pgnumfont\number\pageno\hfil\headlinefont\the\AUTHORHEAD}
\def\righthead{\headlinefont\the\ARTHEAD\hfil\pgnumfont\number\pageno}
\def\firsthead{\headlinefont Baltic Astronomy,~vol.\the\VOLUME,
\the\PAGES,~\the\year .\hfil}
\voffset=2\baselineskip 
%

\newdimen\oldbaselineskip \oldbaselineskip=\baselineskip
\def\test#1{\newlinechar=`@\if!\the#1! \message{#1 not given@}\fi}%
\def\printheader{
  \parindent=0pt
  \null\vskip1.cm
  \test{\TITLE}
  \vbox{\baselineskip=15pt
    \titlefont\the\TITLE
    }
  \vskip8mm plus8mm
  \test{\AUTHOR}
  \authorfont\the\AUTHOR
  \vskip2mm
  \test{\ADDRESS}
  \addressfont\the\ADDRESS
  \vskip2mm
  \test{\SUBMITTED}
  \line{\datefont Received \the\SUBMITTED
    \if!\the\ACCEPTED!\else, accepted \the\ACCEPTED\fi.\hfill}
  \vskip4mm plus4mm
  \vbox{\leftskip=\sumindent\parindent=0pt
    \parskip=5pt
    \absfont Abstract.
    \test{\SUMMARY}
    \sumfont\the\SUMMARY\par
    \absfont Key words:
    \test{\KEYWORDS}
    \sumfont\the\KEYWORDS\par
    }
  \sumfont
  \if!\the\SENDOFF!\else\footnote{}{Send offprint requests to:
 \the\SENDOFF}\fi
  \parindent=2em
  }
%
%
\newdimen\uppergap \newdimen\lowergap
\uppergap=5mm \lowergap=3mm
\newdimen\secind \newdimen\subsecind \newdimen\subsubsecind
\setbox0=\hbox{\secfont 9. }\secind=\wd0
\setbox0=\hbox{\subsecfont 9.9. }\subsecind=\wd0
\setbox0=\hbox{\subsubsecfont 9.9.9. }\subsubsecind=\wd0
\def\section#1{\goodbreak\par\vskip\uppergap
  \noindent\hangindent\secind\hangafter=1\secfont#1
  \vskip\lowergap\mainfont\par\nobreak}
\def\subsection#1{\goodbreak\par\vskip\uppergap
  \noindent\hangindent\subsecind\hangafter=1\subsecfont#1
  \vskip\lowergap\mainfont\par\nobreak}
\def\subsubsection#1{\goodbreak\par\vskip\uppergap
  \noindent\hangindent\subsubsecind\hangafter=1\subsubsecfont#1
  \vskip\lowergap\mainfont\par\nobreak}
%
%
\def\WFigure#1#2#3{\goodbreak\midinsert\vbox{
  \null\centerline{#2}\vskip1.5truemm
  \figheadfont\indent Fig.~#1.\figfont\ #3
  \par\mainfont
  }\endinsert}
%

%

%

%

%
\newdimen\tabind
\setbox0=\hbox{\tabheadfont Table 55.} \tabind=\wd0
\def\Table#1#2{\noindent
  \hangindent\tabind\hangafter=1
  \tabheadfont Table~#1.\tabfont #2
 \par
  \mainfont
  }
%
%
\def\References{\vskip\uppergap
\line{\secfont REFERENCES\hfill}
  \vskip0.8\lowergap
 \petitrm
  }
\def\ref{\goodbreak
\hangindent12pt\hangafter=1
\noindent\ignorespaces}
\def\endref{\egroup}
%
%
\def\byebye{\egroup\par\vfill\supereject\end}
%
%

%
%

\def\utw{\smash{\rlap{\lower5pt\hbox{$\sim$}}}}
\def\udtw{\smash{\rlap{\lower6pt\hbox{$\approx$}}}}


\font\tabfont=cmr9



\def\ddown{\lower2.5ex\hbox}
\def\ddow{\lower1.7ex\hbox}
\def\down{\lower1ex\hbox}
\def\uppp{\raise1ex\hbox}
\def\dnnn{\lower1ex\hbox}
\def\uuppp{\raise2ex\hbox}

\def\ts{\thinspace}
\def\(o-c){$O-C$}


\def\angstr{A\kern-.56em\raise1.9ex\hbox{$\scriptscriptstyle\circ$}$\,$}

\newdimen\free\newdimen\shift
\def\Entry#1#2#3{\par\goodbreak\smallskip%
  \setbox1=\vbox{\advance\hsize by-10mm\parindent=0pt
    \def\\{\par}%
    \it#1. \rm#2}
  \line{\box1\hfill#3}\smallskip
}%
\newdimen\savesize

\def\shiftfigure #1#2#3#4#5{
    \vbox to #2 { \ifodd #5 \rightskip#4 \else\leftskip#4 \fi
                  \null\vfil
                  \figheadfont Fig.~#1.\figfont #3
                  \medskip
                }
                          }

\year1998

\input psfig.sty
\def\ts{\thinspace}

\year 2002
\VOLUME {~11}
\PAGES {341--366}
\pageno=341

\TITLE={KURUCZ MODEL ENERGY DISTRIBUTIONS:\hfil\break  A COMPARISON WITH
REAL STARS. \hfil\break II. METAL-DEFICIENT STARS}

\AUTHOR={V. Strai\v zys$^1$, R. Lazauskait\.e$^{1,2}$ and G.
Valiauga$^1$}

\AUTHORHEAD={V. Strai\v zys, R. Lazauskait\.e, G. Valiauga}

\ARTHEAD={Kurucz model energy distributions. II}

\ADDRESS={$^1$ Institute of Theoretical Physics and Astronomy,
Go\v stauto 12, \hfil\break\null~~~ Vilnius 2600, Lithuania;

$^2$ Department of Theoretical Physics, Vilnius Pedagogical University,
\hfil\break\null~~~ Student\c u 39, Vilnius 2340, Lithuania.}

\SUBMITTED={July 25, 2002}

\SUMMARY={Energy distributions of synthetic spectra for Kurucz model
atmospheres are compared with observed energy distributions of
metal-deficient stars of the blue horizontal-branch (BHB), F--G--K
subdwarf (SD) and G--K giant (MDG) types.  The best coincidence of the
synthetic and observed energy curves is found for BHB stars.  The
largest differences are found in the ultraviolet wavelengths for
subdwarfs and cool MDGs.  The influence of errors of effective
temperatures, gravities and metallicities is estimated.}

\KEYWORDS={stars:  fundamental parameters, spectral energy distribution
-- stars: horizontal-branch, subdwarfs, metal-deficient giants}

\printheader

\section{1. INTRODUCTION}

Synthetic energy distributions of the Kurucz models (Kurucz 1995) are
frequently used for a selection of optimum positions of passbands for
multicolor photometric systems and for calculation of parameters of
photometric systems or of transformation equations between color indices
of different photometric systems.  However, the use of synthetic spectra
for these tasks is justified only in the case if the synthetic spectra
of model atmospheres coincide sufficiently well with energy
distributions of real stars having the same physical parameters
(effective temperatures, gravities and metallicities).  \vskip0.5mm

In our first paper (Strai\v zys, Liubertas \& Valiauga 1997) we have
compared energy distributions of the Kurucz model atmospheres with the
mean energy distributions of stars of various spectral types of solar
metallicity published by Strai\v zys \& Sviderskien\.e (1972) and
revised by one of the authors.  In the present paper we compare the
Kurucz synthetic spectra of metal-deficient models with spectral energy
distributions of real stars -- blue horizontal-branch (hereafter, BHB)
stars, F--G subdwarfs (hereafter, SD) and G--K metal-deficient giants
(hereafter, MDG) published by Sviderskien\.e (1992).

\section{2. INTRINSIC ENERGY DISTRIBUTION CURVES OF STARS}

Energy distribution functions of 52 stars -- 9 blue
horizontal-branch stars, 11 extreme F--G--K subdwarfs, 10 mild F--G
subdwarfs, 14 mild metal-deficient G--K giants and subgiants and 6
extreme metal-deficient G--K giants -- were published by Sviderskien\.e
(1992).  The wavelength dependence of energy fluxes for each of these
stars was determined by averaging the data from all sources found in the
literature until 1990.  For the majority of stars the flux data cover
the range from 300 to 1050 nm.  For some stars observed with the IUE
orbiting observatory, the fluxes are available down to 150--170 nm.  The
energy distribution curves are given in the absolute-relative form, i.e.
they all are normalized to 100 at 550 nm.  The mean energy distribution
of each star is given in the intensities per unit wavelength interval.
Each value given in the catalog is the intensity integral in the
interval $\Delta\lambda$ = 5 nm centered on a given wavelength.  The
stars affected by interstellar extinction, are dereddened.  The accuracy
of the observed energy distributions will be estimated in the last
section of this paper.  \vskip0.5mm

In the catalog of Sviderskien\.e (1992), Table 4 on pages 44--48 has a
typographic error -- the numbers of stars are given in wrong positions.
The right order of HD numbers in the heading of the Table should be:  HD
127665, 135722, 216763, 217906, 219615 and 2857 (instead of 217906,
219615, 2857, 127665, 135722 and 216763).

\section{3. A COMPARISON OF THE MODEL AND THE OBSERVED\hfil\break
SPECTRA}

For the present investigation we have selected only the stars with
extreme metal-deficiency:  6 blue horizontal-branch stars, 11 extreme
subdwarfs and 6 extreme metal-deficient giants.  These stars are listed
in Table 1. \vskip0.5mm

Effective temperatures, gravities and metallicities for most of these
stars are taken from Cenarro et al.  (2001) where these parameters are
reduced to a unique system based on high-dispersion spectral analysis.
According to that paper, the estimated errors of the parameters are
$\pm$100 K for $T_{\rm e}$, $\pm$0.2 dex for $\log g$ and $\pm$0.15
dex for [Fe/H].  For some stars the parameters are taken from other
sources listed in the Notes to Table 1.
\vskip0.5mm

Sviderskien\.e (1992) has excluded the interstellar extinction effect
on energy distribution curves for some giants which were found to be
slightly reddened. We have made a new search of interstellar reddenings
for the stars of Table 1. The values of color excesses, given in the
Table, were used to deredden energy distributions of the stars.
Intrinsic energy distributions $I_0 ({\lambda})$ were calculated from
the observed $I\ts ({\lambda})$ by the equation
$$
I_0 (\lambda) = {I (\lambda)\over {\tau^{x}(\lambda)}}, \eqno(1)
$$
where $\tau (\lambda)$is the transmittance function of the unit mass
of interstellar dust and $x$ is the amount of unit masses. The values of
$x$ for different color excesses were calculated by the equation
$$
x = {E_{B-V}\over E_{B-V}^{(x=1)}}. \eqno(2)
$$
Here $E_{B-V}^{(x=1)}$ for stars of various spectral types are
calculated from the standard interstellar extinction law $\tau(\lambda)$
given in Table 3 of the Strai\v zys (1992) monograph using the following
equations:
$$
A_m =~-2.5\log
{\int {I\ts (\lambda)R_m(\lambda)\tau^x(\lambda)d\lambda}\over
\int{I\ts (\lambda)R_m(\lambda) d\lambda}} .
\eqno(3)$$
and
$$ E_{m_1-m_2} = A_{m_1} - A_{m_2}. \eqno(4)$$
Here $I\ts (\lambda)$ are the intrinsic energy distribution functions of
stars, $R_m(\lambda)$ are the response functions of the $B$ and $V$
passbands, $m_1$ and $m_2$ are $B$ and $V$ magnitudes. Mean energy
distribution functions for stars of various spectral classes were taken
from Strai\v zys \& Sviderskien\.e (1972) with the slightly revised
ultraviolet. The values of $E_{B-V}^{(x=1)}$ due to the band-width
effect slightly depend on the intrinsic energy distribution of the star:
for O--B--A stars (including BHB stars) it is close to 1.00 and for
the G5--K5 giants (including MDG stars) it is $\sim$0.85.

\vskip0.5mm

For the stars, which were dereddened by Sviderskien\.e, only
differential reddening due to different color excess value was taken
into account.  \vskip0.5mm

\topinsert

\Table{1}{ A list of the stars whose spectral energy distribution data
are compared with the Kurucz models.}
\vskip-3mm
$$\vbox{\tabskip 33pt minus 33pt\tabfont
\halign to \hsize {\hfil #  & \hfil #  & \hfil # & \hfil # & \hfil # \cr
\noalign{\medskip\hrule\medskip}
HD, BD  & {\itl T}$_{\rm e}$ & log\ts {\itl g} & [Fe/H] & {\itl
E}$_{B-V}$ \cr
\noalign{\medskip\hrule\medskip}
\noalign{\tabfont {BHB stars}}
\noalign{\smallskip}
2857    & 7563  &  2.67  &  --1.60  &  0.02  \cr
60778\rlap{$^a$} & 8690  &  3.30  &  --0.50  &  0.01  \cr
74721\rlap{$^b$} & 8640  &  3.55  &  --1.48  &  0.01  \cr
86986\rlap{$^b$} & 7850  &  3.10  &  --1.81  &  0.02  \cr
109995  & 8034  &  2.98  &  --1.55  &  0.00  \cr
161817  & 7639  &  2.96  &  --0.95  &  0.01  \cr
\noalign{\smallskip}
\noalign{\tabfont {F--G--K subdwarfs}}
\noalign{\smallskip}
19445   & 5918  &  4.35  &  --2.05  &  0.00  \cr
25329   & 4787  &  4.58  &  --1.72  &  0.00  \cr
64090\rlap{$^c$}  &  5446  &  4.45  &  --1.76  &  0.00  \cr
84937  &  6228  &  4.01  &  --2.17  &  0.00  \cr
94028  &  5941  &  4.21  &  --1.49  &  0.00  \cr
103095 &  5025  &  4.56  &  --1.36  &  0.00  \cr
140283 &  5687  &  3.55  &  --2.53  &  0.01  \cr
188510 &  5490  &  4.69  &  --1.59  &  0.00  \cr
219617 &  5878  &  4.04  &  --1.39  &  0.00  \cr
+26 2606\rlap{$^d$} & 6146 &  4.23  &  --2.29  &  0.01  \cr
+17 4708 & 6005 &  4.01  &  --1.74  &  0.01  \cr
\noalign{\smallskip}
\noalign{\tabfont {G--K metal-deficient giants}}
\noalign{\smallskip}
2665   &  5013  &  2.35  &  --1.96  &  0.06  \cr
6755   &  5102  &  2.40  &  --1.41  &  0.04  \cr
88609  &  4513  &  1.26  &  --2.64  &  0.02  \cr
122563 &  4566  &  1.12  &  --2.63  &  0.01  \cr
165195 &  4471  &  1.11  &  --2.15  &  0.15  \cr
221170\rlap{$^c$}  &  4465  &  1.04  &  --2.10  &  0.03  \cr
\noalign{\medskip\hrule\medskip}
}}$$

\noindent\tabfont{NOTES:}

\tabfont{
\noindent $^a$ HD~60778: parameters are from Danford \& Lea (1981).

\noindent $^b$ HD~74721 and HD 86986: parameters are from Adelman \&
Philip (1994).

\noindent $^c$ HD~64090 and HD~221170:  parameters are from Soubiran et
al.  (1998).

\noindent $^d$ BD~+26~2606: parameters are from Gratton et al. (1996).

}

\endinsert

Figures 1--23 compare the energy distribution curves of Kurucz and
Sviderskien\.e for the metal-deficient stars of Table 1. The
temperatures of the models are rounded to the nearest number multiple of
10.  For each star two Kurucz energy curves are plotted with the
temperatures $\pm$200 K from the right value.  The space between these
two curves is shown in grey.  The synthetic curves for the specific
$T_{\rm e}$, $\log g$ and [Fe/H] are obtained by linear interpolation
between the grid curves given by Kurucz.  The following conclusions can
be made for different types of stars.  \vskip0.5mm

(1) For the BHB stars the coincidence of energy distributions
is relatively good.  However some stars show considerable differences in
some spectral ranges.  For example, the model of HD 60778 (Figure~2)
is stronger by 10--20\% in the ultraviolet and violet and fainter by
$\sim$5\% in the red and infrared.  This difference can be corrected by
taking a lower value of $T_{\rm e}$ of the star.  The model energy
curve of HD 161817 (Figure~6) is also higher by $\sim$10\% in the
ultraviolet and violet, but no difference is seen in long wavelengths.
\vskip0.5mm

(2) In case of the extreme subdwarfs the coincidence is relatively good
in the whole spectral range (from the ultraviolet to the infrared) only
for HD 25329 (Figure 8) and BD+26 2606 (Figure 16).  For the remaining
nine subdwarfs the model curves, corresponding to the $T_{\rm e}$ values
given in Table 1, are too low by 10--15\% in the blue, violet and
ultraviolet ranges.  These differences cannot be explained by a $T_{\rm
e}$ error, since in the red and infrared the model and the observed
curves are close to each other.  Only for HD 19445 (Figure 7) an
increase in the temperature of the star by 200 K would force the
observed and the model curves to be in sufficient agreement.
\vskip0.5mm

(3) For the extreme metal-deficient giants the coincidence of the
observed and model curves in the red and infrared is also good.  For HD
2665 (Figure 18) and HD 165195 (Figure 22) the model curve is good also
in the violet and ultraviolet.  However, for the other four stars there
are systematic differences in the short wavelength part of the spectrum:
there is a general weakening of model spectra in the violet and
ultraviolet for HD 6755, HD 88609, HD 122563 and HD 221170.  The upper
limit of the shadowed area in short wavelengths almost coincides with
the real energy curve for all these four stars.  This might indicate
that $T_{\rm e}$ accepted for the stars is by $\sim$200 K too low.
However, acceptance of higher temperature will lead to a disagreement of
energy curves in the red and infrared.  \vskip0.5mm

For the evaluation of the effect of gravity uncertainties, we have
chosen three stars, representing BHB, SD and MDG stars.  The value of
$\log g$ was varied by $\pm$0.3 dex from the true value, keeping the
$T_{\rm e}$ and [Fe/H] values constant.  Model energy curves for these
stars are compared with the observed energy curves in Figures 24--26.
It is evident that the gravity effects are of minor importance and they
cannot help to explain the differences seen in Figs. 1--23.  The same is
true in the case of [Fe/H] variations by $\pm$0.3 dex (Figures 27--29).

\section{4. A COMPARISON OF SYNTHETIC COLOR INDICES}

For a quantitative evaluation of the differences between the model and
the observed energy distributions, we calculated and intercompared
synthetic color indices in the {\it Vilnius} photometric system and the
{\it UBV} system of the metal-deficient stars from Table 1 and the stars
of luminosities V and III of solar chemical composition (hereafter,
normal stars).  The method of synthetic photometry used is described by
Strai\v zys (1996).  Energy distributions for normal stars were taken
from Strai\v zys \& Sviderskien\.e (1972) (with the corrected
ultraviolet, as described in Paper I) and for metal-deficient stars --
from Sviderskien\.e (1992).  The response functions of the {\it Vilnius}
and {\it UBV} systems were taken from Strai\v zys (1992).

The results are shown in Figs. 30--37, where the differences $\Delta$ of
corresponding color indices (``model'' minus ``observed'') are plotted
against the effective temperatures.  The temperatures for
metal-deficient stars are taken from Table 1 and for the normal stars --
from Paper I. \vskip0.5mm

Synthetic color indices reflect the differences seen in energy
distributions.  For stars of solar chemical composition (dots and
crosses) the scatter of stars gradually increases with decreasing
temperature.  However, for stars with $T_{\rm e}$$>$4500 K the
deviations $\Delta$ do not exceed $\pm$0.2 mag even in case of color
indices containing the ultraviolet magnitudes.  The deviations of
subdwarfs are also within these limits.  The largest deviations
(plus 0.3--0.4 mag) are observed for the three coolest metal-deficient
giants with the temperatures near 4500 K. It is probable, that in the
ultraviolet these stars are really brighter than their Kurucz models.

\section{5. DISCUSSION AND CONCLUSIONS}

The comparison of energy distributions of real stars and their
models, described in the previous sections, is a difficult problem for
interpretation since there are several causes affecting the accuracy of
the results.  They are:  (1) accidental and systematic errors of the
observed energy distributions, (2) systematic errors in the
account of opacity sources in model stellar atmospheres, and (3) errors
of the accepted effective temperatures of the stars.  Let us discuss
these error sources separately. \vskip0.5mm

When we plot energy distributions of a star, determined by different
authors, and normalize all of them at one wavelength, say at 550 nm,
then we find a considerable scatter of individual determinations at both
ends of the wavelength interval covered.  If we limit ourselves with the
traditional ground-based data, covering the wavelengths between 320 nm
and 1 $\mu$m, then the usual scatter of the energy flux values for
G--K--M type stars is:  $\pm$5--10\% at 350 nm and $\pm$5\% at 1 $\mu$m
(Strai\v zys \& Sviderskien\.e 1972, Sviderskien\.e 1988).  The flux
values obtained as averages from the results of several authors should
be more accurate.  \vskip0.5mm

A good test of the accuracy of the average spectral energy curves is a
comparison of color indices obtained from these curves by synthetic
photometry and color indices of the same stars observed
photoelectrically with sufficiently high accuracy ($\pm$0.01 mag).  Such
a comparison in the passbands of the {\it Vilnius} photometric system
has been done both for normal stars of various luminosities
(Sviderskien\.e 1988) and for metal-deficient stars (Sviderskien\.e
1992).  In both cases the stars exhibit differences between the
ultraviolet synthetic and observed color indices within $\pm$0.10 mag.
Consequently, a considerable part of the differences seen in Figures
1--37 can be explained by the errors in observed energy distributions.
\vskip0.5mm

It is strange enough that the differences of the observed and the model
energy distributions give an evidence that at least part of
metal-deficient stars (subdwarfs and giants) exhibit overestimation of
opacity in stellar atmospheres.  It is more usual that synthetic spectra
of model atmospheres demonstrate a ``missing opacity".  On the other
hand, some model stars exhibit a perfect coincidence of their spectral
energy distributions with observations.  If the opacity were a problem,
then all stars of a given type would behave in the same way.
\vskip0.5mm

The errors of the accepted effective temperatures is the third
possibility.  Although Cenarro et al.  (2001) estimate their
temperatures to be accurate within $\pm$100 K, it is not excluded that
in some cases these errors may be twice larger.  The vertical bars in
Figures 35--37 show how the temperature errors of $\pm$200 K affect the
plotted differences of color indices.  It seems that these error bars
are capable to explain the scatter of most stars in these figures,
except of the stars cooler than $\sim$4500 K. \vskip0.5mm

Thus we conclude that the present investigation gives evidence that
spectral energy distributions of the Kurucz metal-deficient models of
the effective temperatures $>$4500 K do not contradict the observed
energy distributions of real stars belonging to the blue horizontal
branch, extreme subdwarfs and extreme metal-deficient \hbox{giants,} if
we take into account the errors of the observational data and of the
effective temperatures.  Cooler models probably exhibit some systematic
differences in the ultraviolet and violet wavelengths.

\vskip2mm

ACKNOWLEDGMENT.  We are grateful to A. G. Davis Philip for his important
comments.

\References

\ref Adelman S. J., Philip A. G. D. 1994, MNRAS, 269, 579

\ref Cenarro A. J., Gorgas J., Cardiel N., Pedraz S., Peletier R. F.,
Vazdekis A. 2001, MNRAS, 326, 981

\ref Danford S. C., Lea S. M. 1981, AJ, 86, 1909

\ref Gratton R. G., Carretta E., Castelli F. 1996, A\&A, 314, 191

\ref Kurucz R. L. 1995, personal communication

\ref Soubiran C., Katz D., Cayrel R. 1998, A\&AS, 133, 221

\ref Strai\v zys V. 1992, Multicolor Stellar Photometry, Pachart
Publishing House, Tucson, Arizona

\ref Strai\v zys V. 1996, Baltic Astronomy, 5, 459

\ref Strai\v zys V., Liubertas R., Valiauga G. 1997, Baltic Astronomy,
6, 601 (Paper I)

\ref Strai\v zys V., Sviderskien\.e Z. 1972, Bull. Vilnius Obs., No. 35,
3

\ref Sviderskien\.e Z. 1988, Bull. Vilnius Obs., No. 80, 3

\ref Sviderskien\.e Z. 1992, Bull. Vilnius Obs., No. 86, 3

\vfil\eject

\WFigure{1}{\psfig{figure=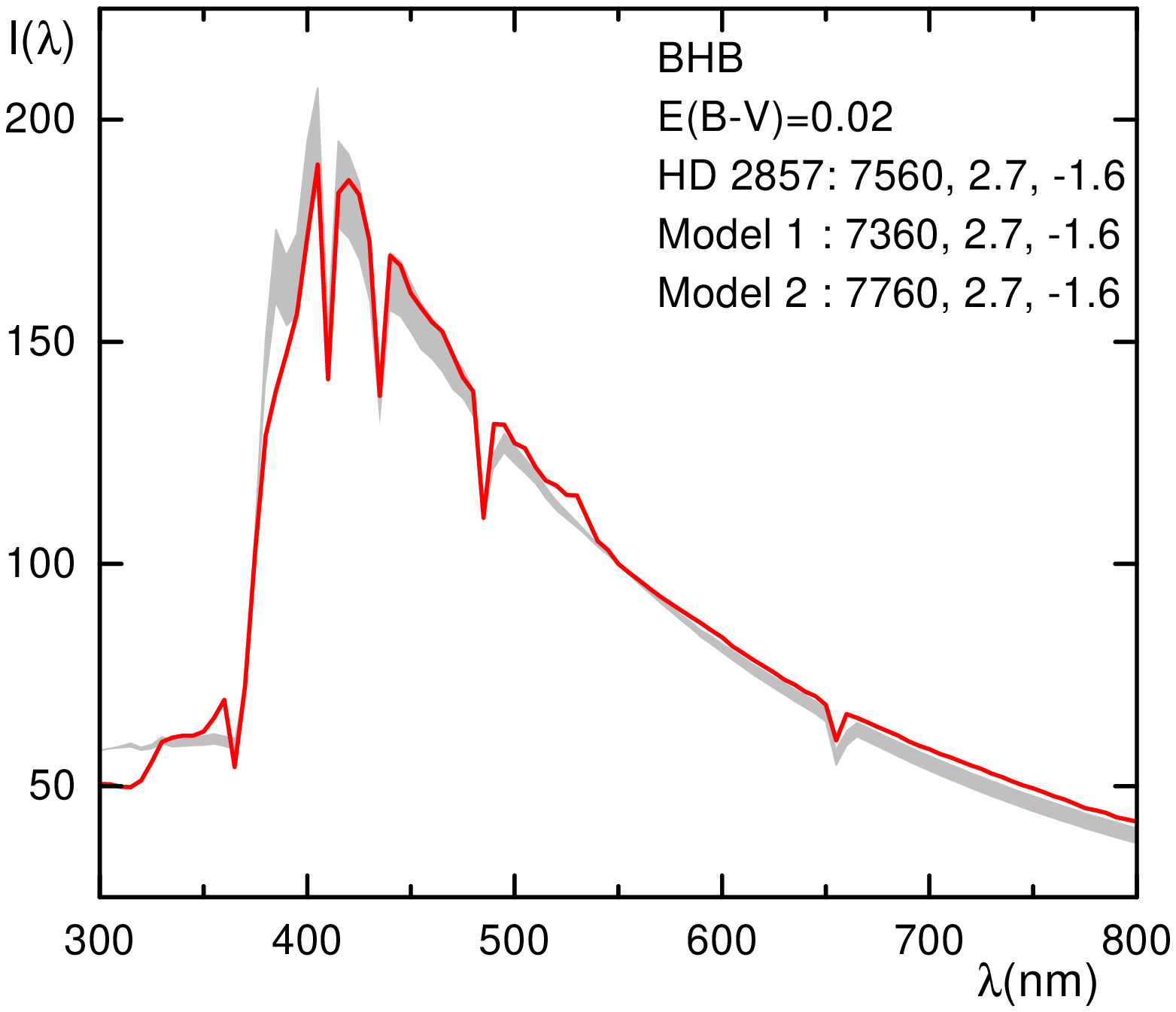,width=9.0truecm,angle=0,clip=}}
{ Spectral energy distribution of HD 2857 (BHB star). The effective
temperature of the model is changed by $\pm$200 K.}

\WFigure{2}{\psfig{figure=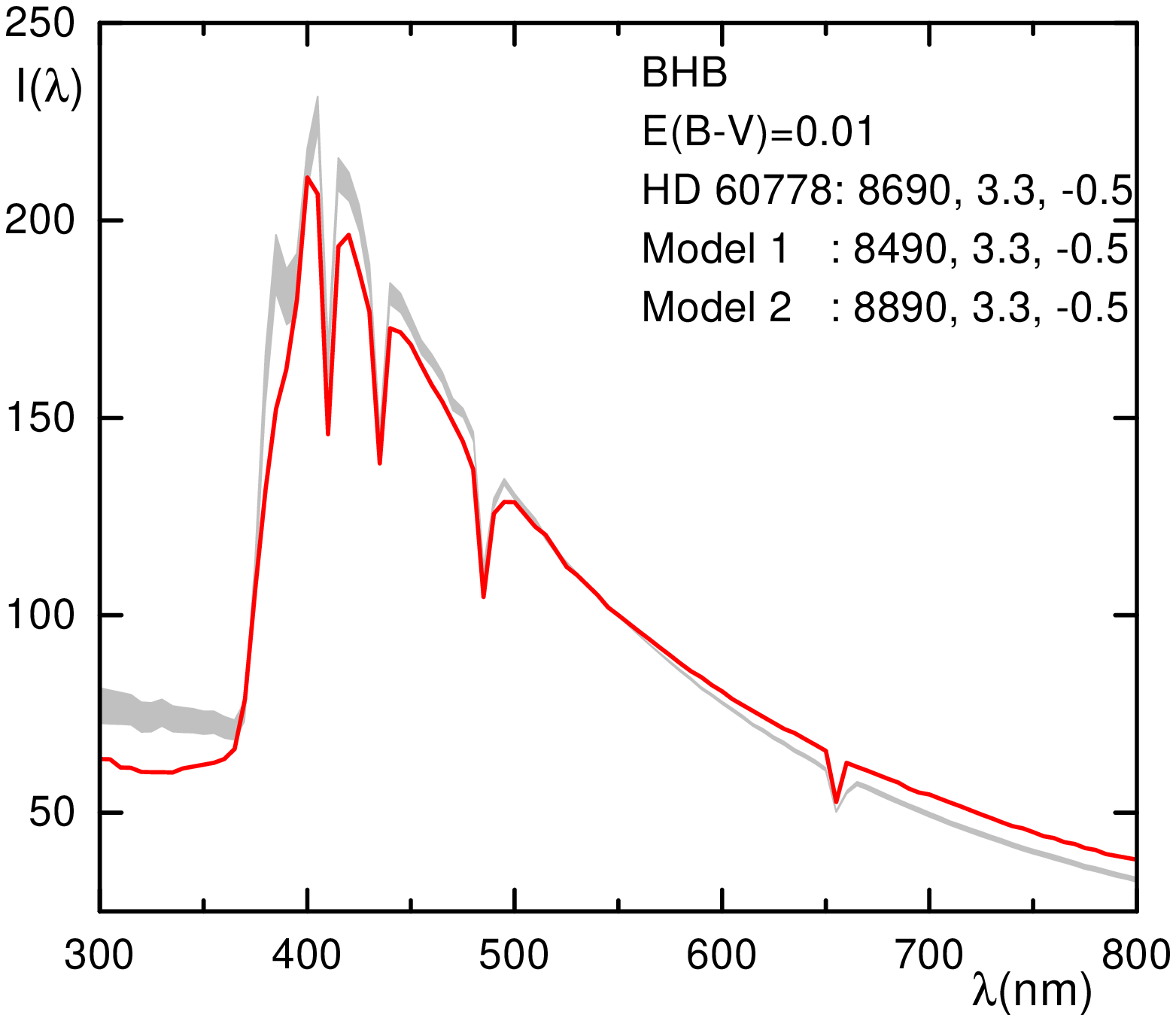,width=9.0truecm,angle=0,clip=}}
{ Spectral energy distribution of HD 60778 (BHB star). The effective
temperature of the model is changed by $\pm$200 K.}

\WFigure{3}{\psfig{figure=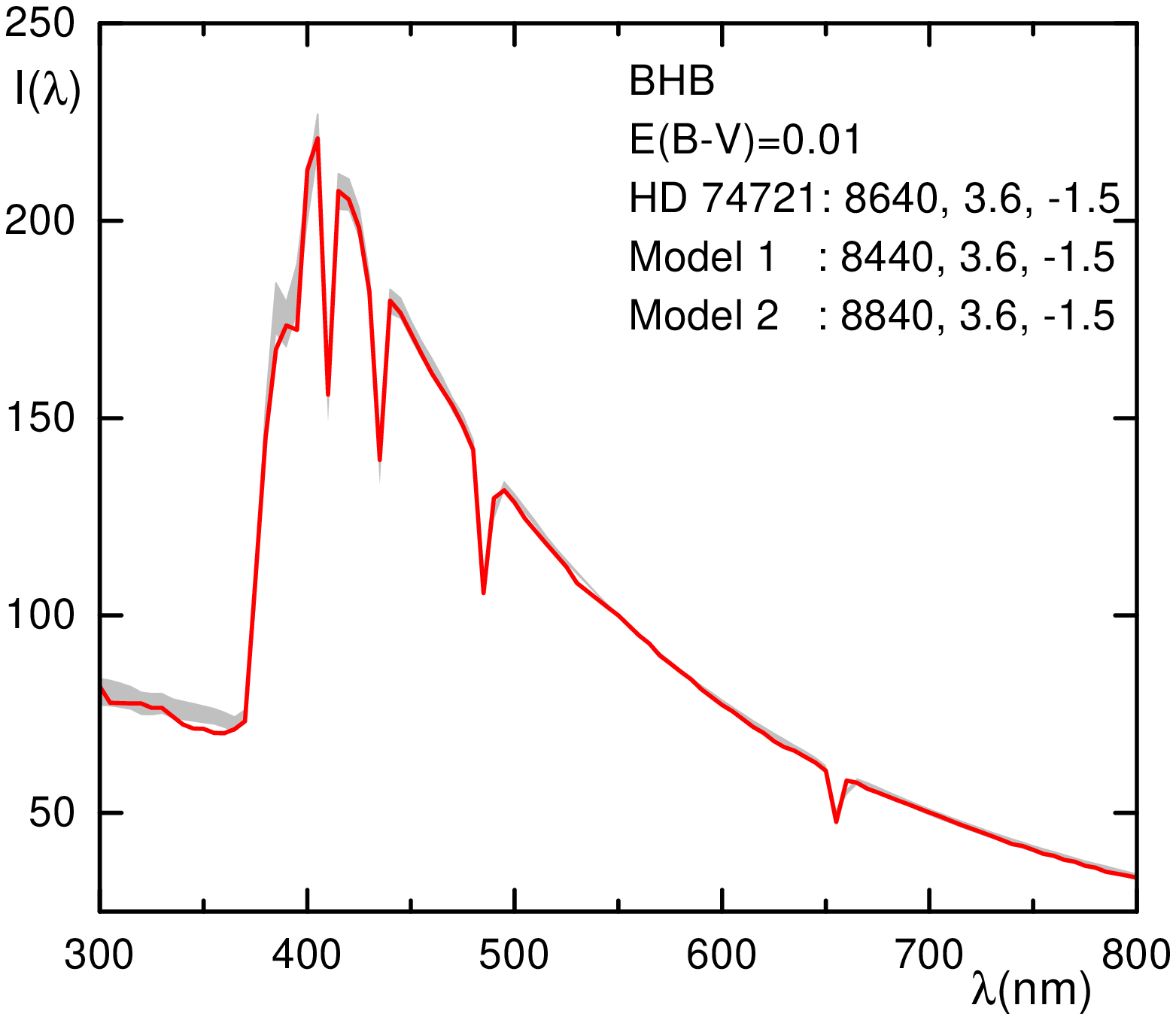,width=9.0truecm,angle=0,clip=}}
{ Spectral energy distribution of HD 74721 (BHB star). The effective
temperature of the model is changed by $\pm$200 K.}

\WFigure{4}{\psfig{figure=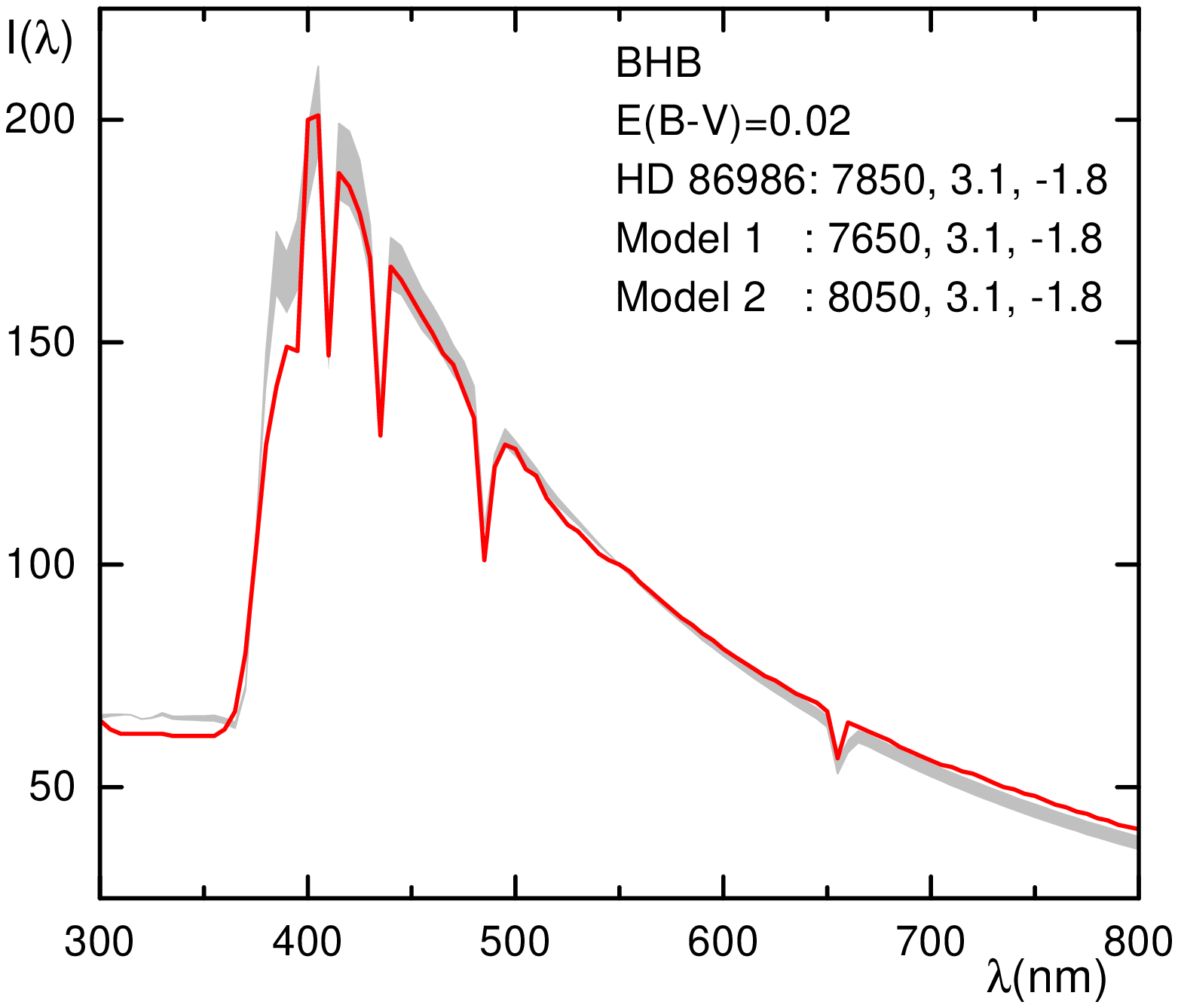,width=9.0truecm,angle=0,clip=}}
{ Spectral energy distribution of HD 86986 (BHB star). The effective
temperature of the model is changed by $\pm$200 K.}

\WFigure{5}{\psfig{figure=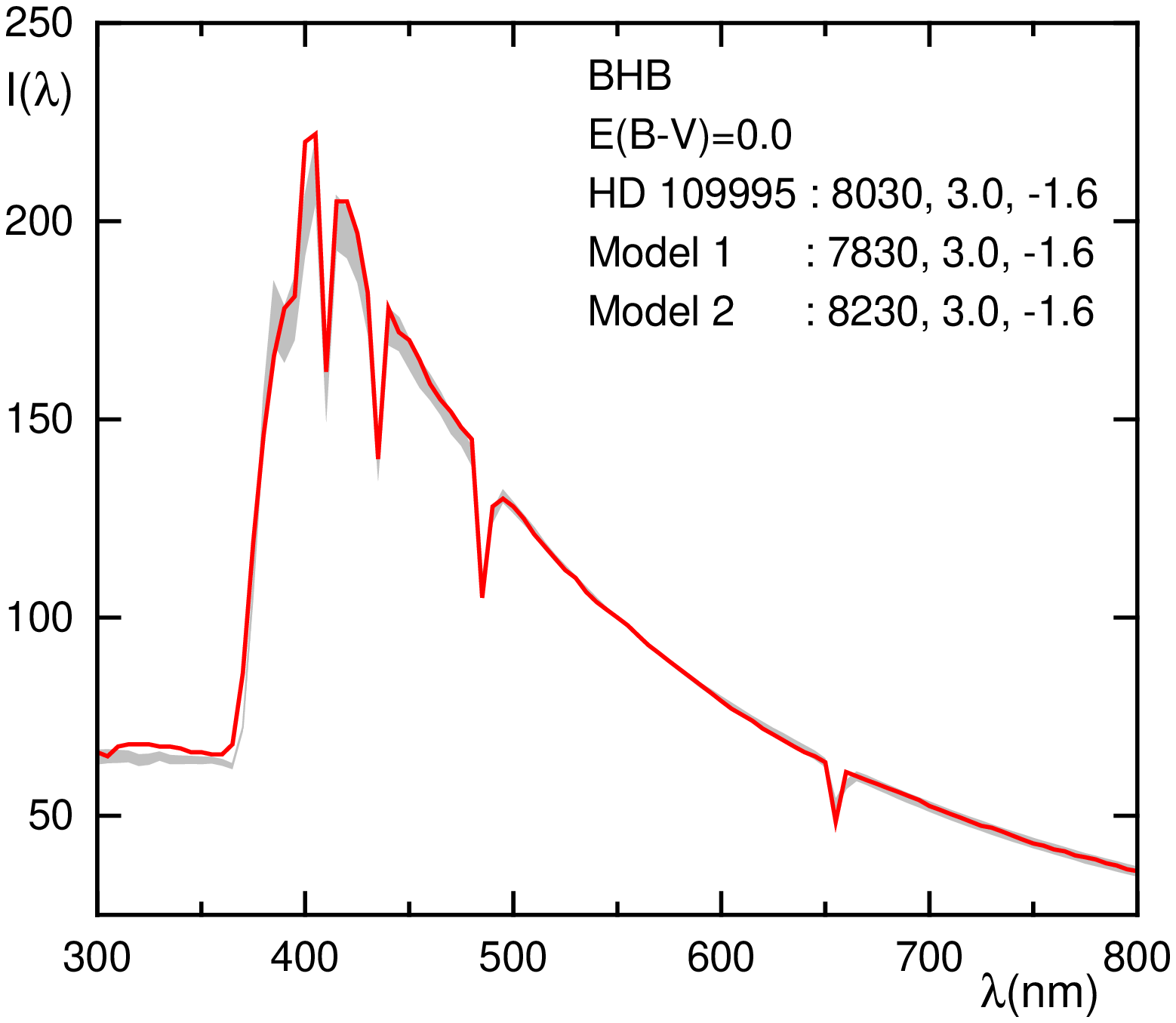,width=9.0truecm,angle=0,clip=}}
{ Spectral energy distribution of HD 109995 (BHB star). The effective
temperature of the model is changed by $\pm$200 K.}

\WFigure{6}{\psfig{figure=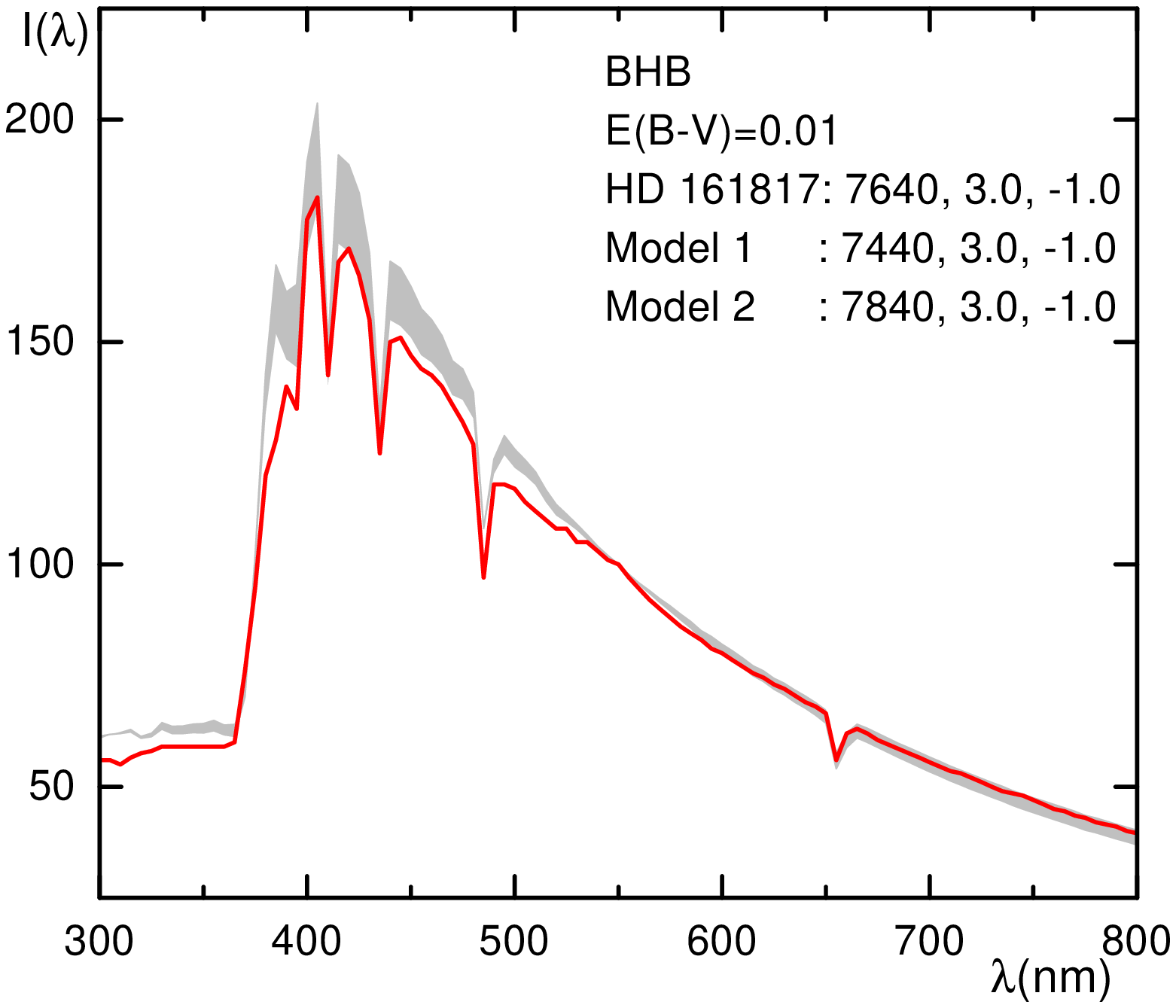,width=9.0truecm,angle=0,clip=}}
{ Spectral energy distribution of HD 161817 (BHB star). The effective
temperature of the model is changed by $\pm$200 K.}

\WFigure{7}{\psfig{figure=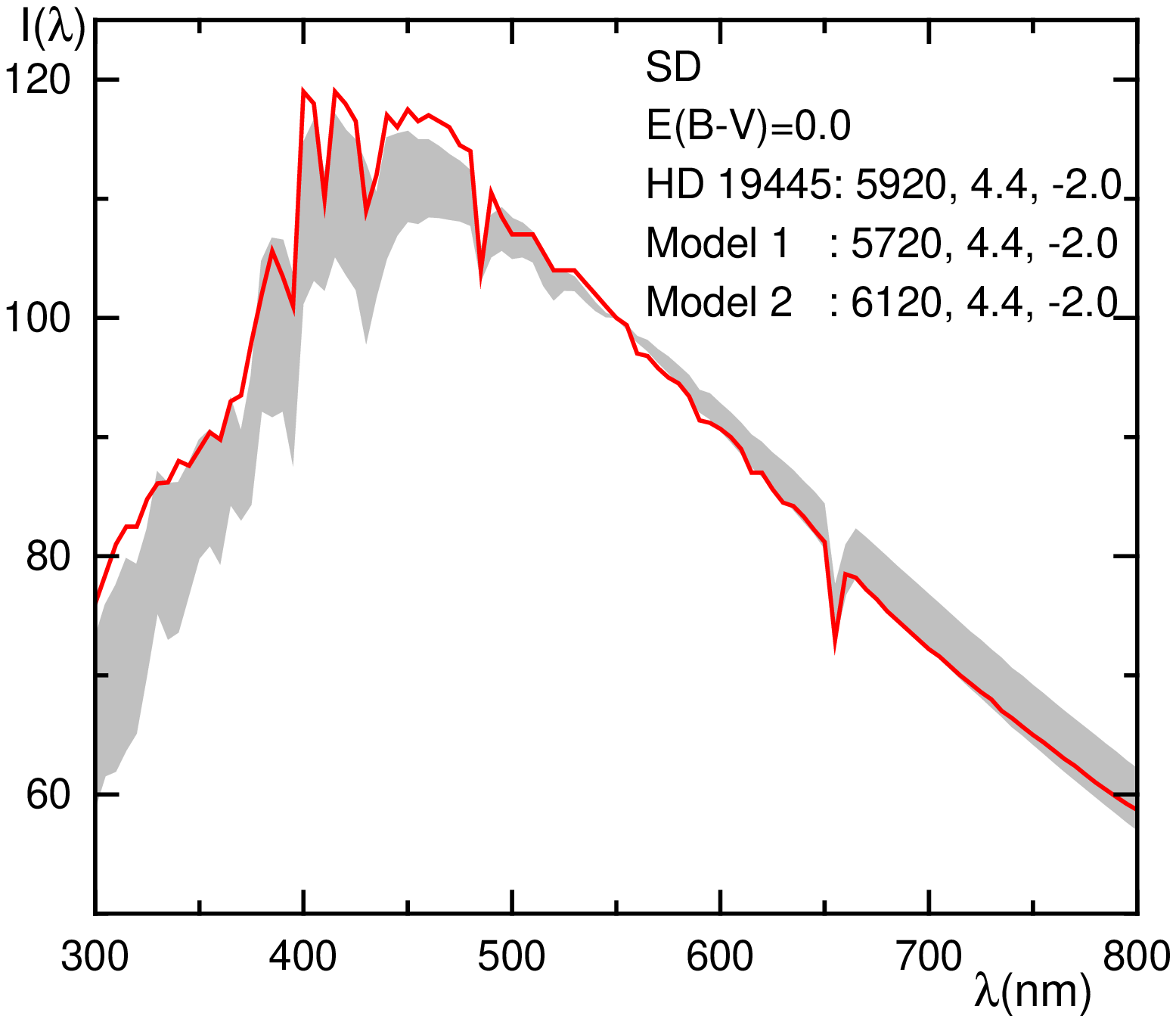,width=9.0truecm,angle=0,clip=}}
{ Spectral energy distribution of HD 19445 (subdwarf). The effective
temperature of the model is changed by $\pm$200 K.}

\WFigure{8}{\psfig{figure=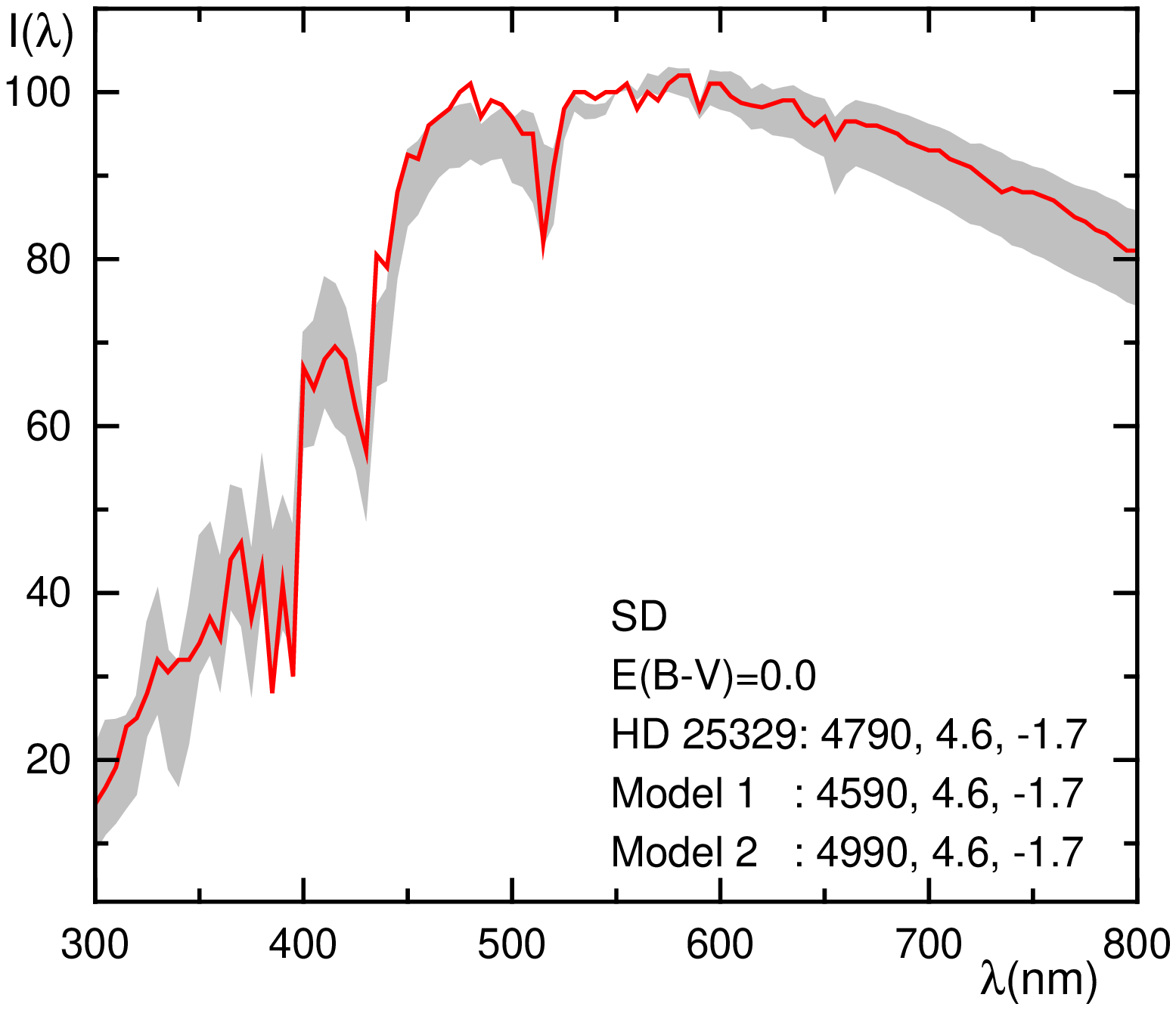,width=9.0truecm,angle=0,clip=}}
{ Spectral energy distribution of HD 25329 (subdwarf). The effective
temperature of the model is changed by $\pm$200 K.}

\WFigure{9}{\psfig{figure=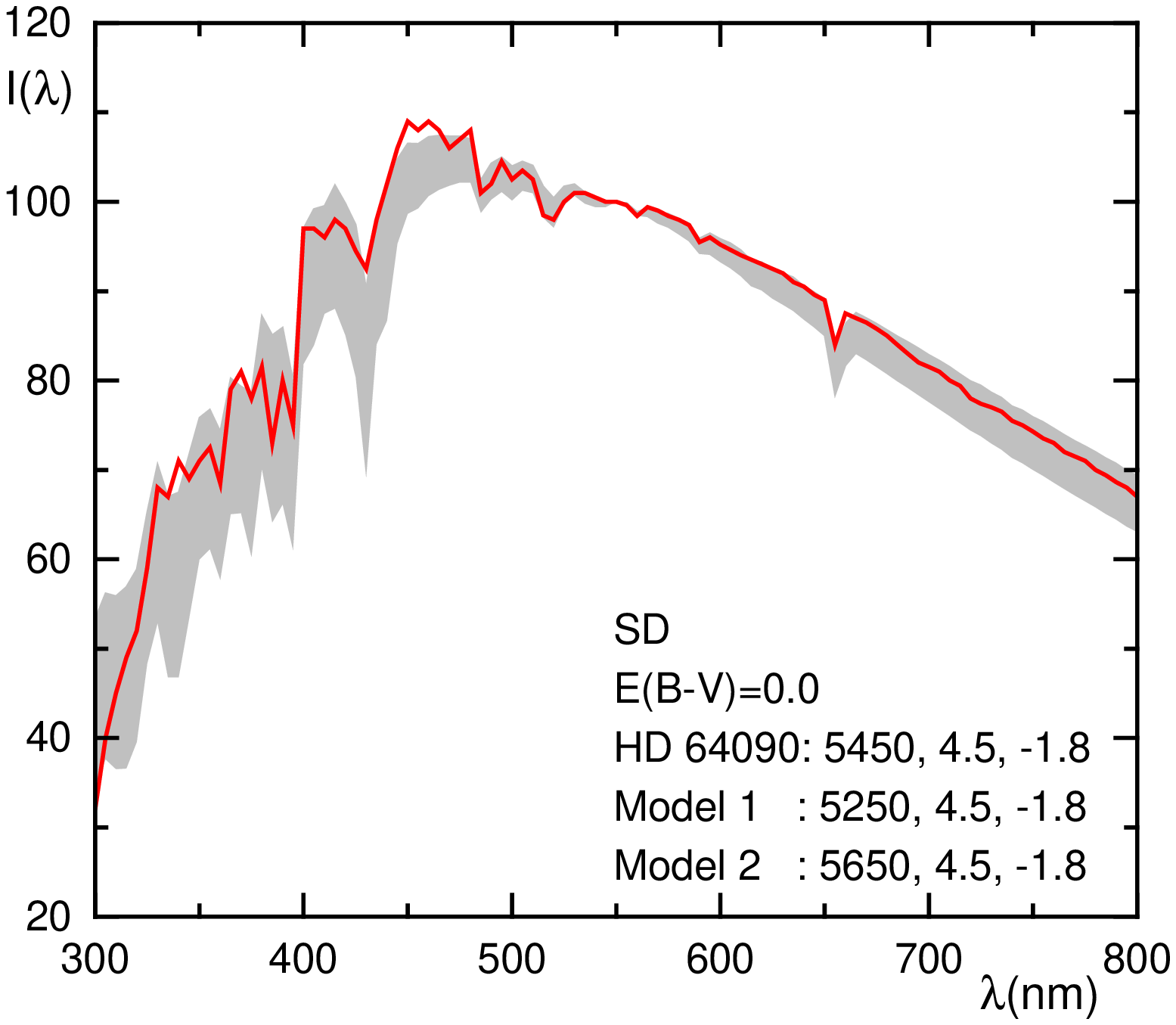,width=9.0truecm,angle=0,clip=}}
{ Spectral energy distribution of HD 64090 (subdwarf). The effective
temperature of the model is changed by $\pm$200 K.}

\WFigure{10}{\psfig{figure=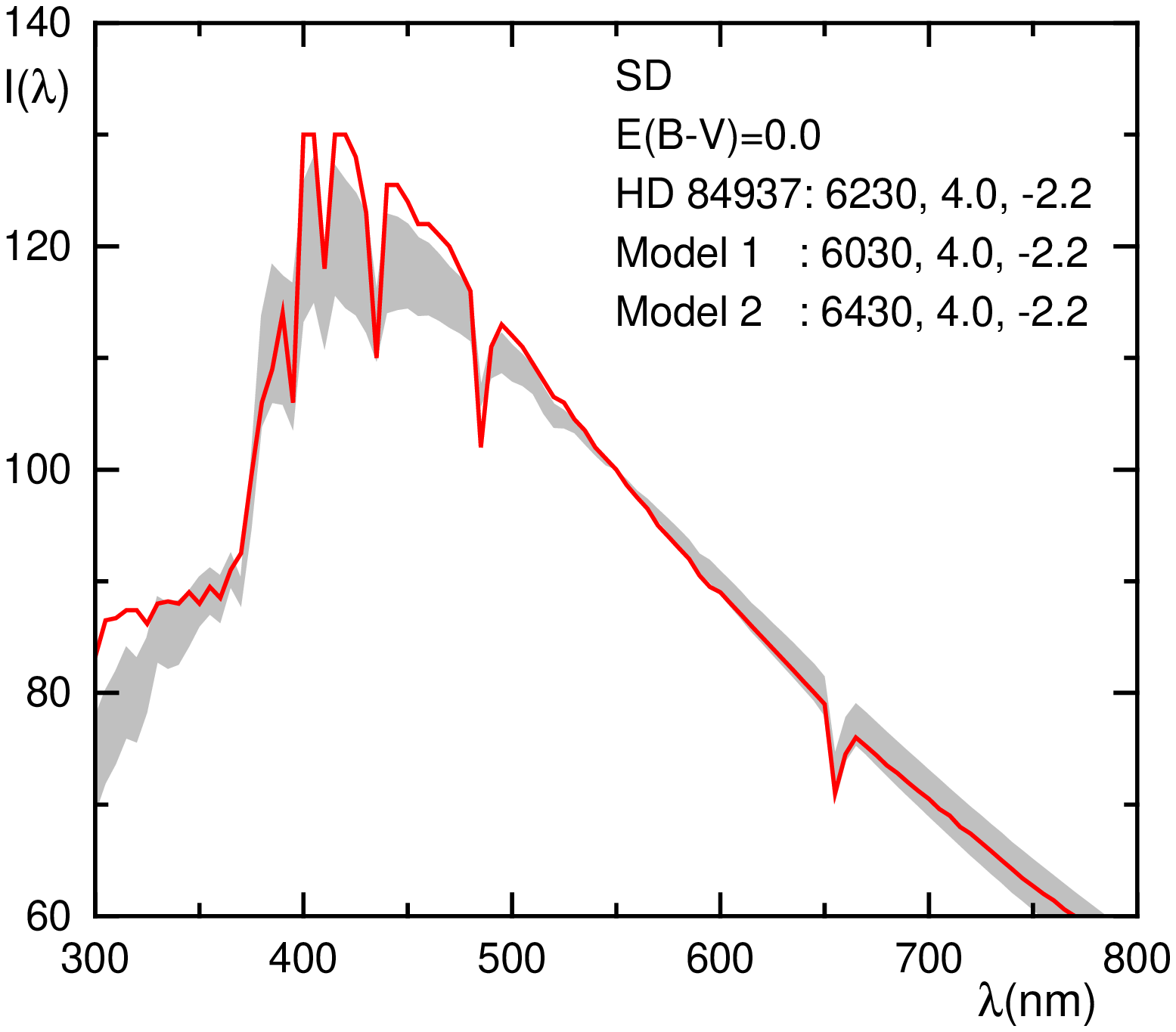,width=9.0truecm,angle=0,clip=}}
{ Spectral energy distribution of HD 84937 (subdwarf). The effective
temperature of the model is changed by $\pm$200 K.}

\WFigure{11}{\psfig{figure=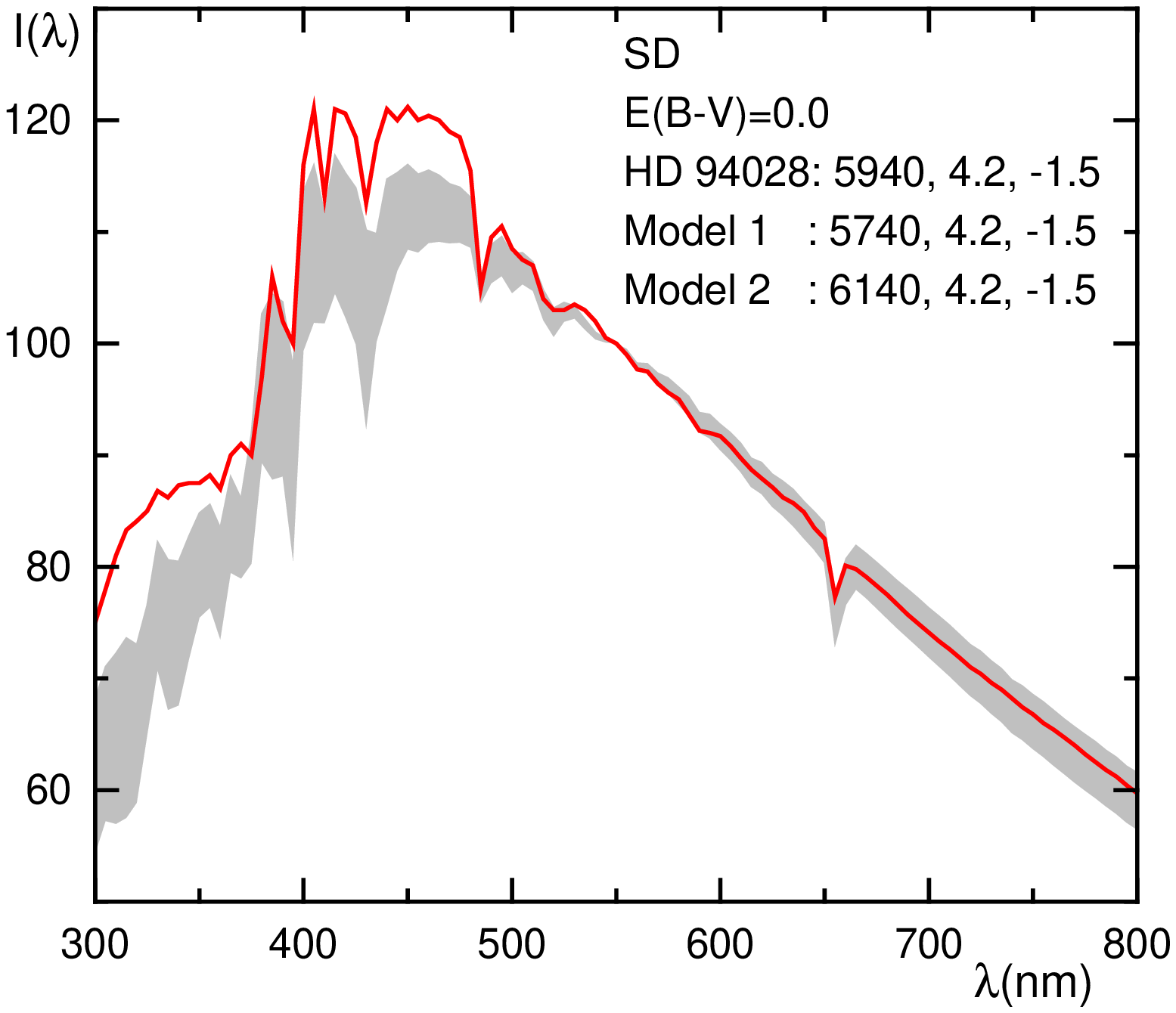,width=9.0truecm,angle=0,clip=}}
{ Spectral energy distribution of HD 94028 (subdwarf). The effective
temperature of the model is changed by $\pm$200 K.}

\WFigure{12}{\psfig{figure=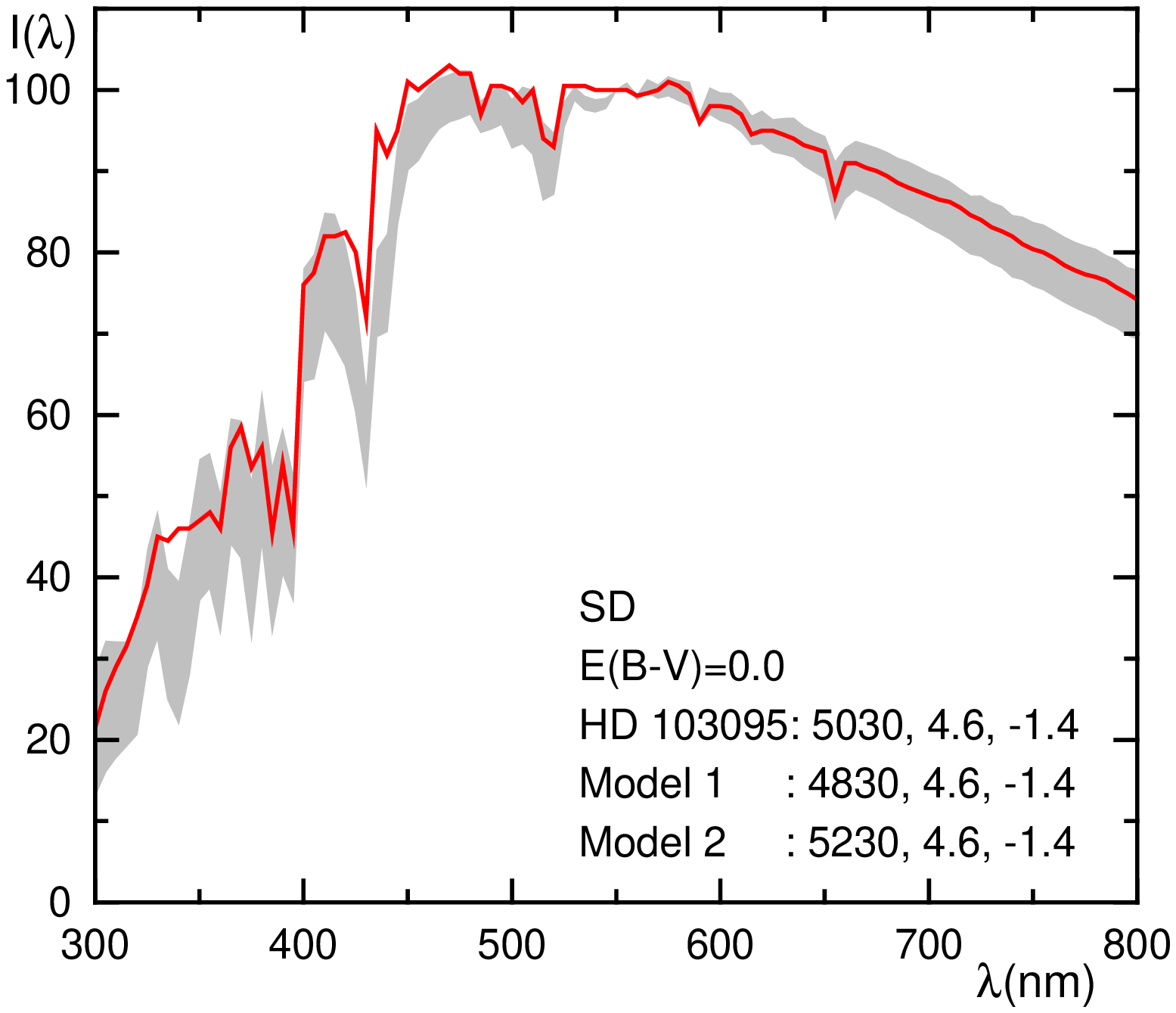,width=9.0truecm,angle=0,clip=}}
{ Spectral energy distribution of HD 103095 (subdwarf). The effective
temperature of the model is changed by $\pm$200 K.}

\WFigure{13}{\psfig{figure=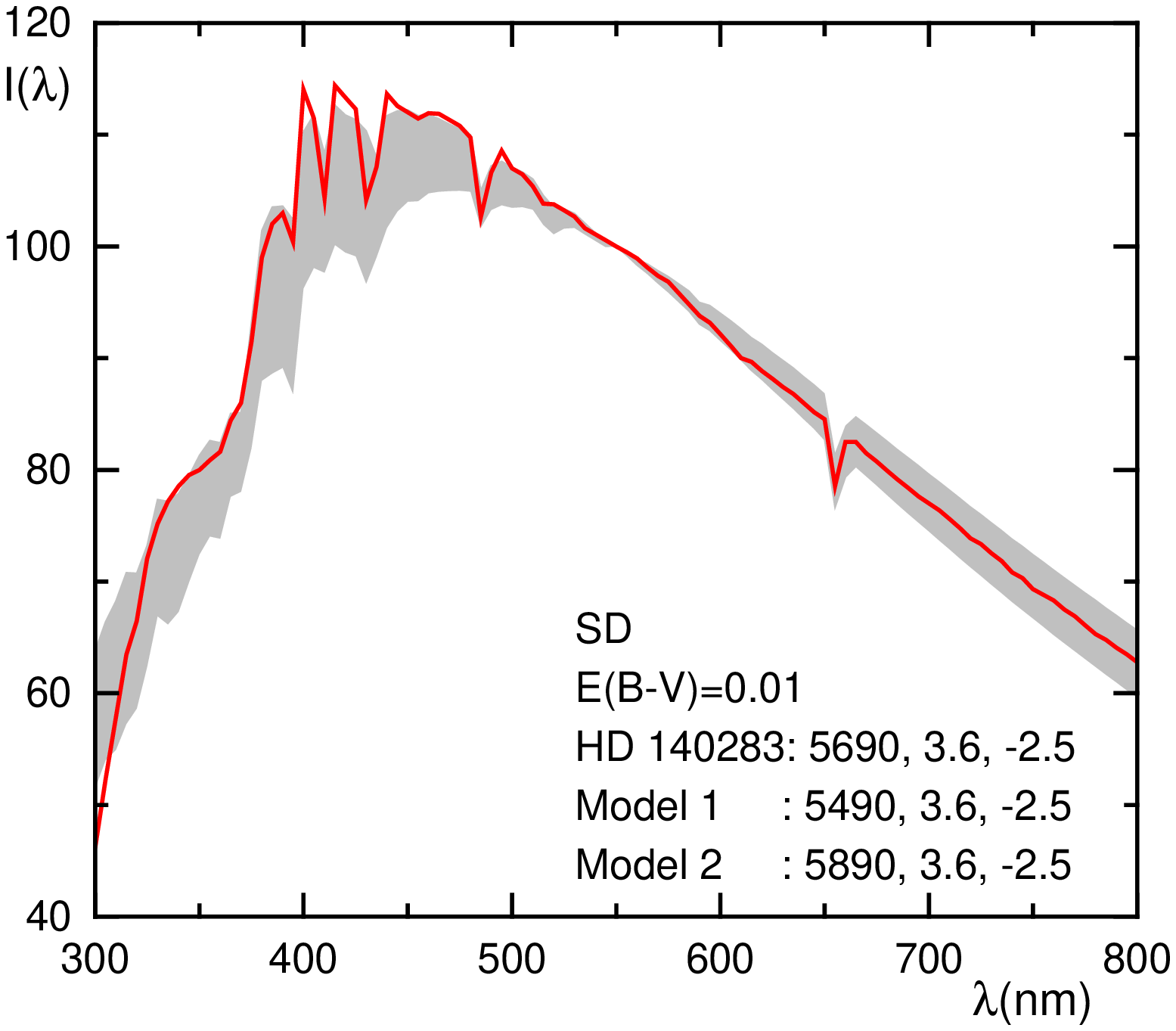,width=9.0truecm,angle=0,clip=}}
{ Spectral energy distribution of HD 140283 (subdwarf). The effective
temperature of the model is changed by $\pm$200 K.}

\WFigure{14}{\psfig{figure=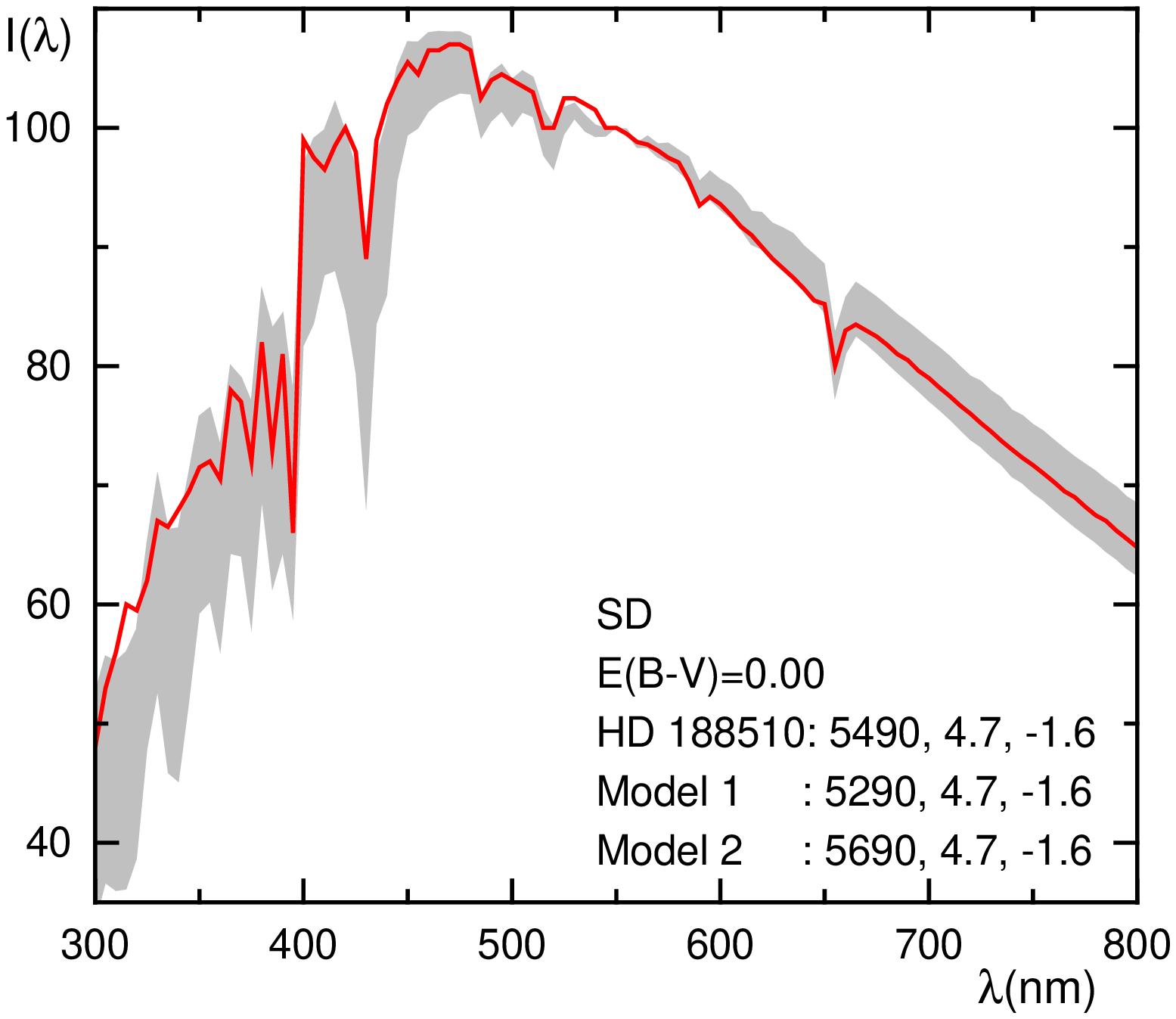,width=9.0truecm,angle=0,clip=}}
{ Spectral energy distribution of HD 188510 (subdwarf). The effective
temperature of the model is changed by $\pm$200 K.}

\WFigure{15}{\psfig{figure=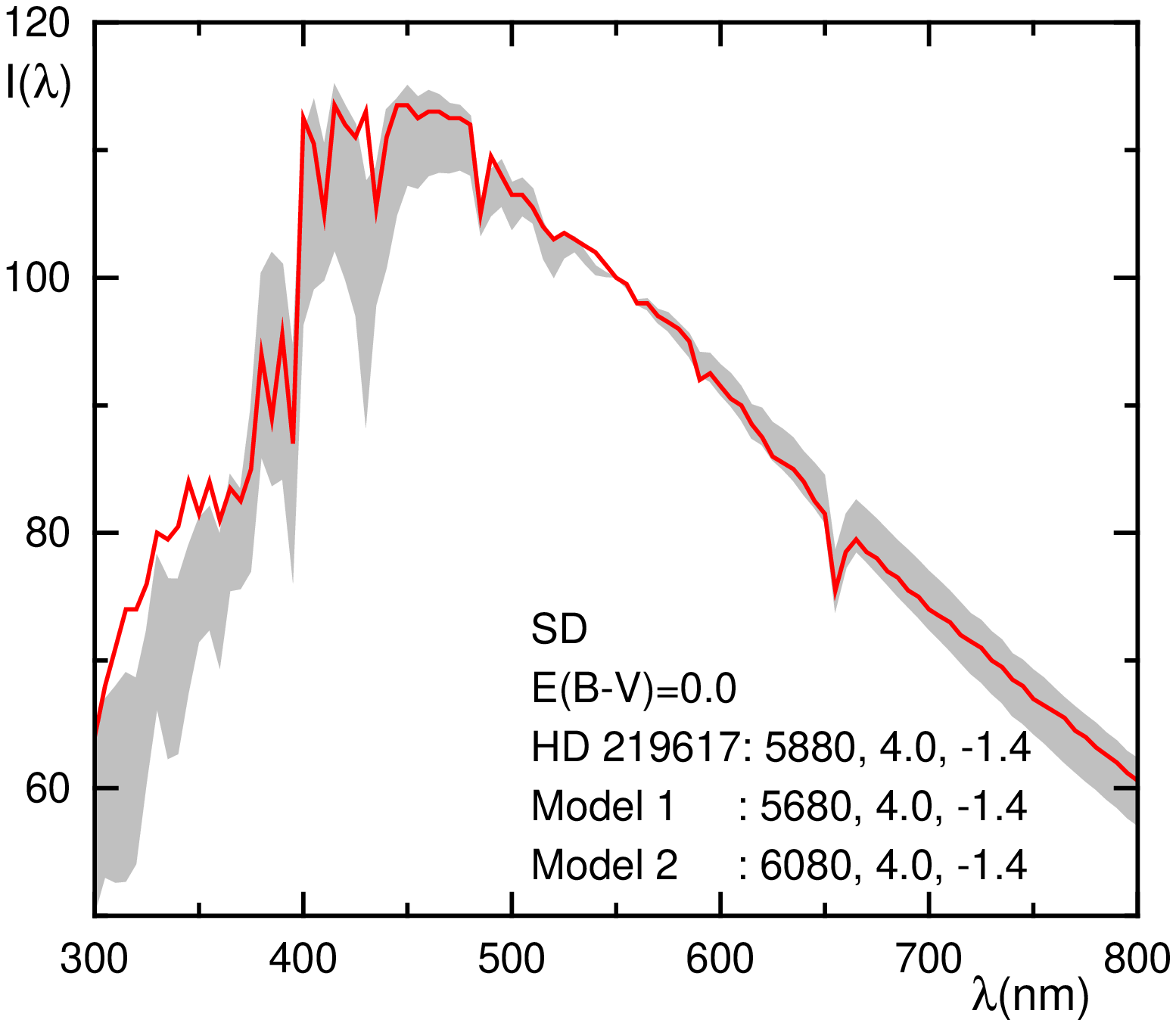,width=9.0truecm,angle=0,clip=}}
{ Spectral energy distribution of HD 219617 (subdwarf). The effective
temperature of the model is changed by $\pm$200 K.}

\WFigure{16}{\psfig{figure=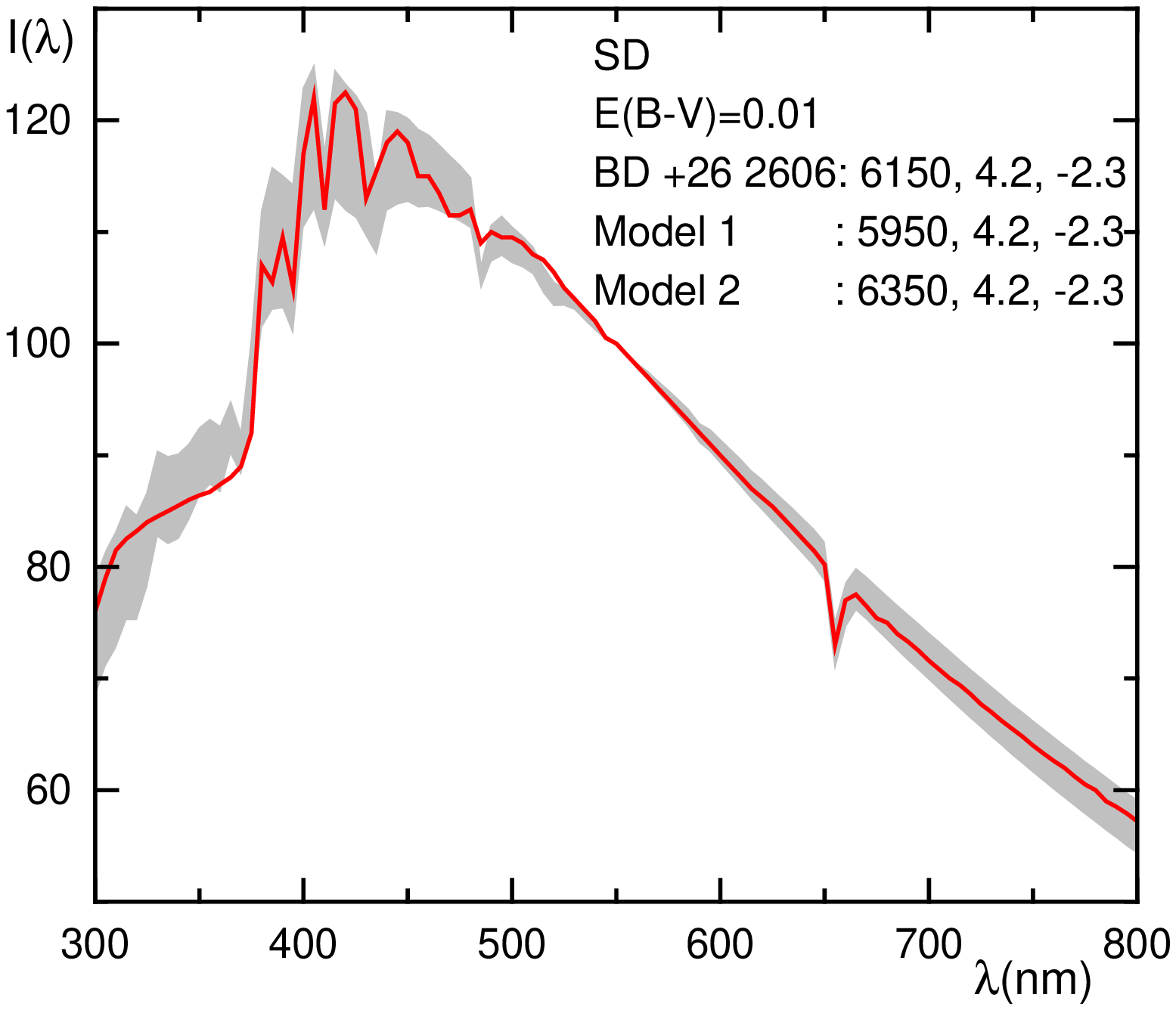,width=9.0truecm,angle=0,clip=}}
{ Spectral energy distribution of BD+26 2606 (subdwarf). The effective
temperature of the model is changed by $\pm$200 K.}

\WFigure{17}{\psfig{figure=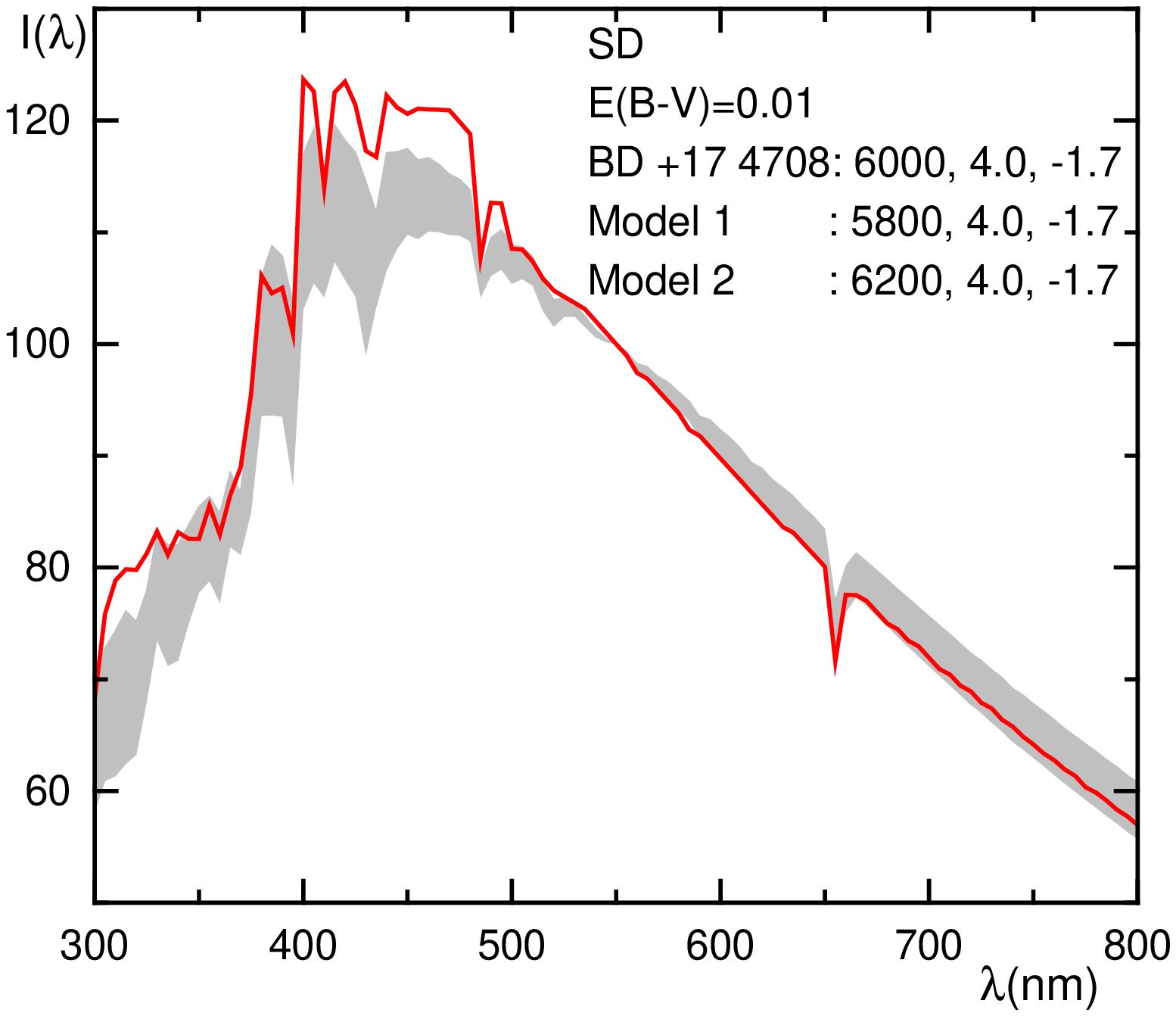,width=9.0truecm,angle=0,clip=}}
{ Spectral energy distribution of BD+17 4708 (subdwarf). The effective
temperature of the model is changed by $\pm$200 K.}

\WFigure{18}{\psfig{figure=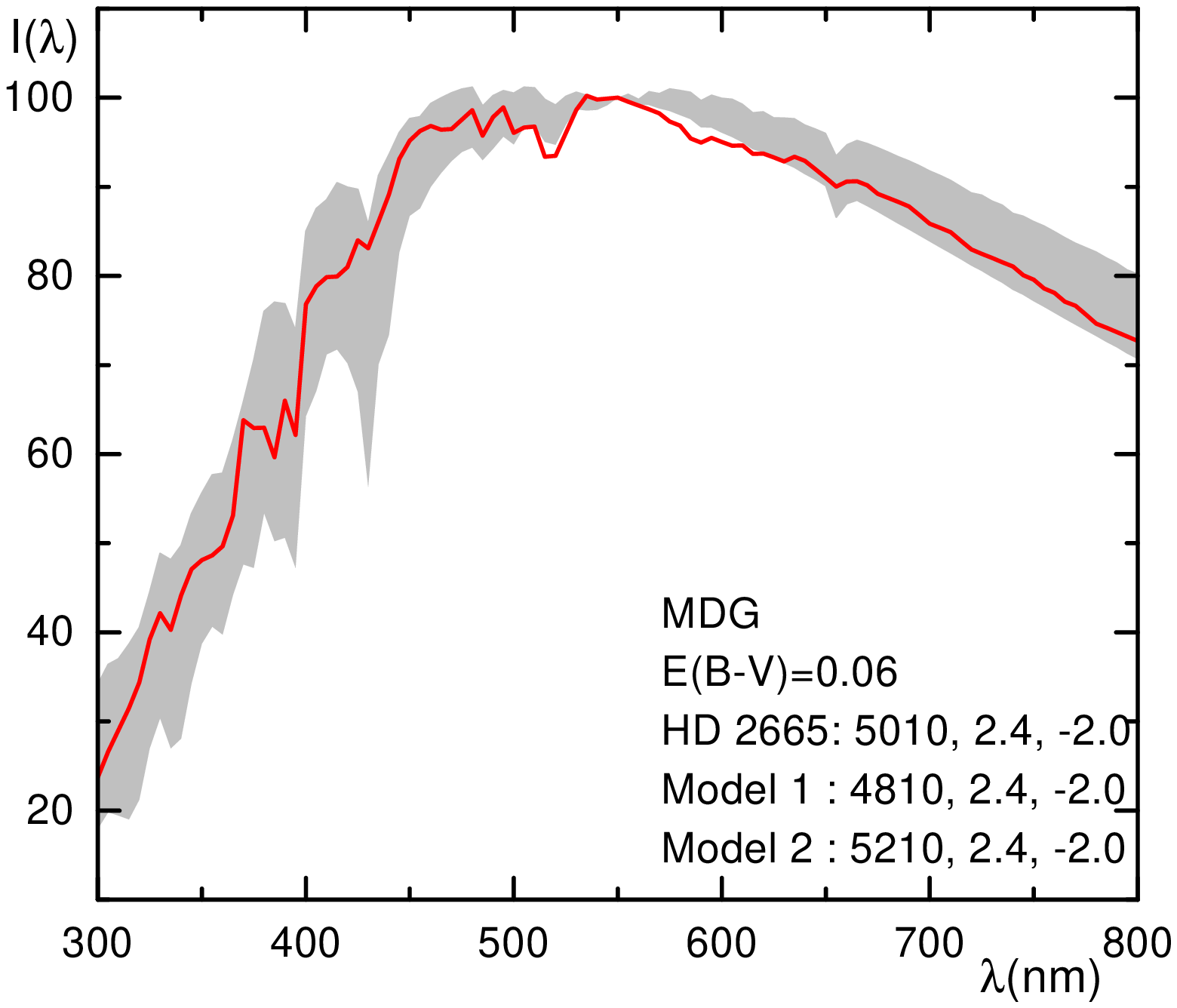,width=9.0truecm,angle=0,clip=}}
{ Spectral energy distribution of HD 2665 (MD giant). The effective
temperature of the model is changed by $\pm$200 K.}

\WFigure{19}{\psfig{figure=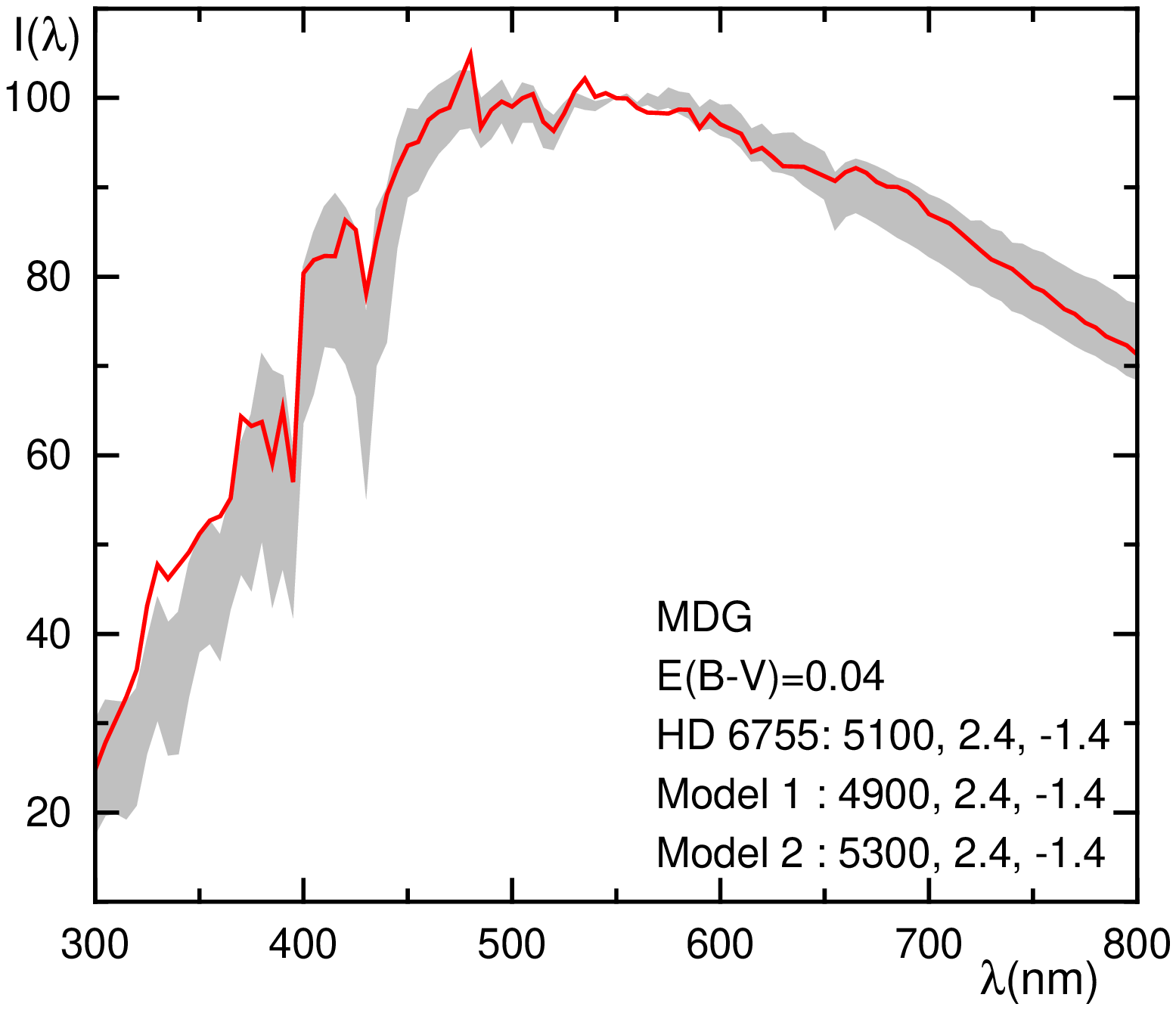,width=9.0truecm,angle=0,clip=}}
{ Spectral energy distribution of HD 6755 (MD giant). The effective
temperature of the model is changed by $\pm$200 K.}

\WFigure{20}{\psfig{figure=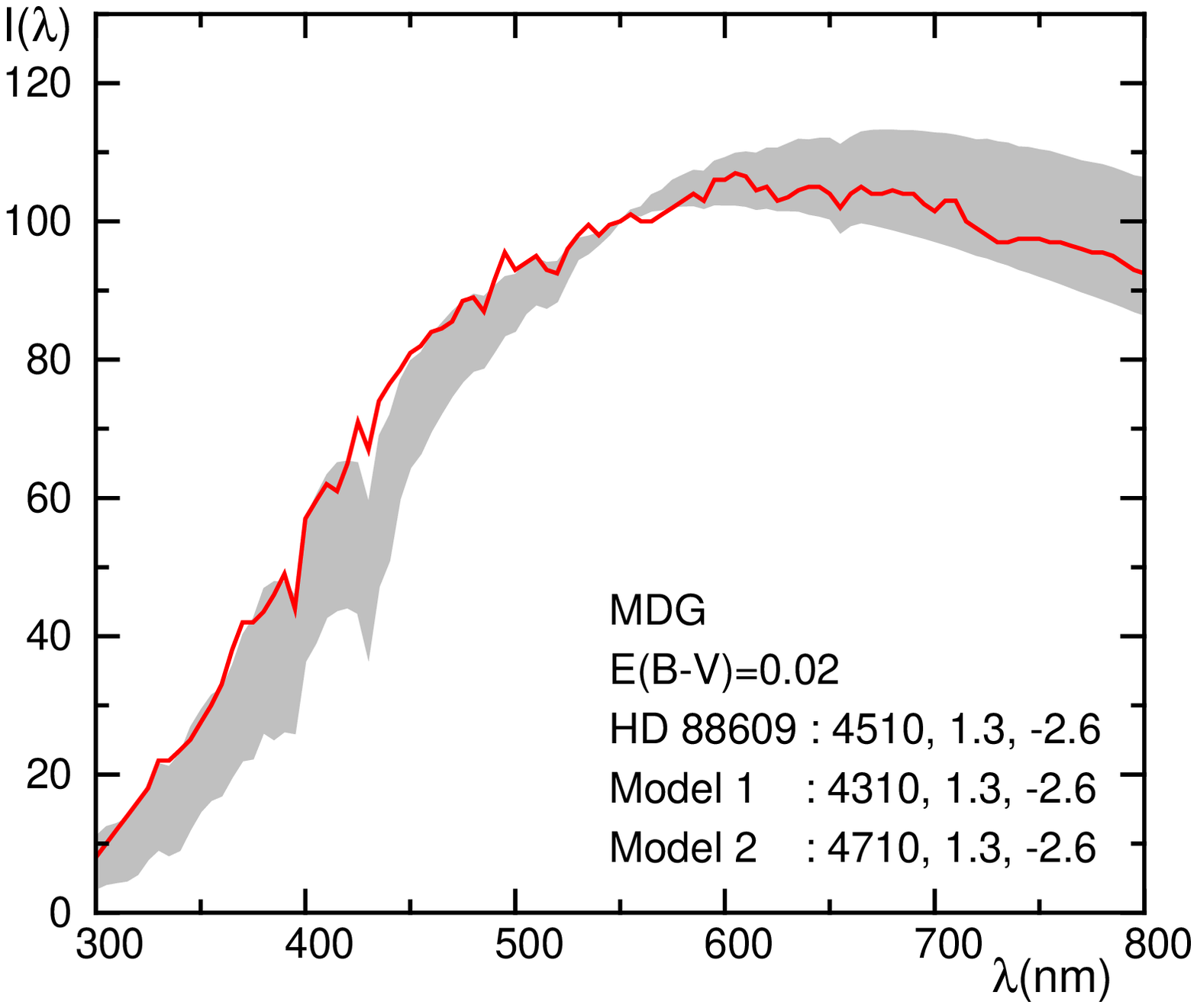,width=9.0truecm,angle=0,clip=}}
{ Spectral energy distribution of HD 88609 (MD giant). The effective
temperature of the model is changed by $\pm$200 K.}

\WFigure{21}{\psfig{figure=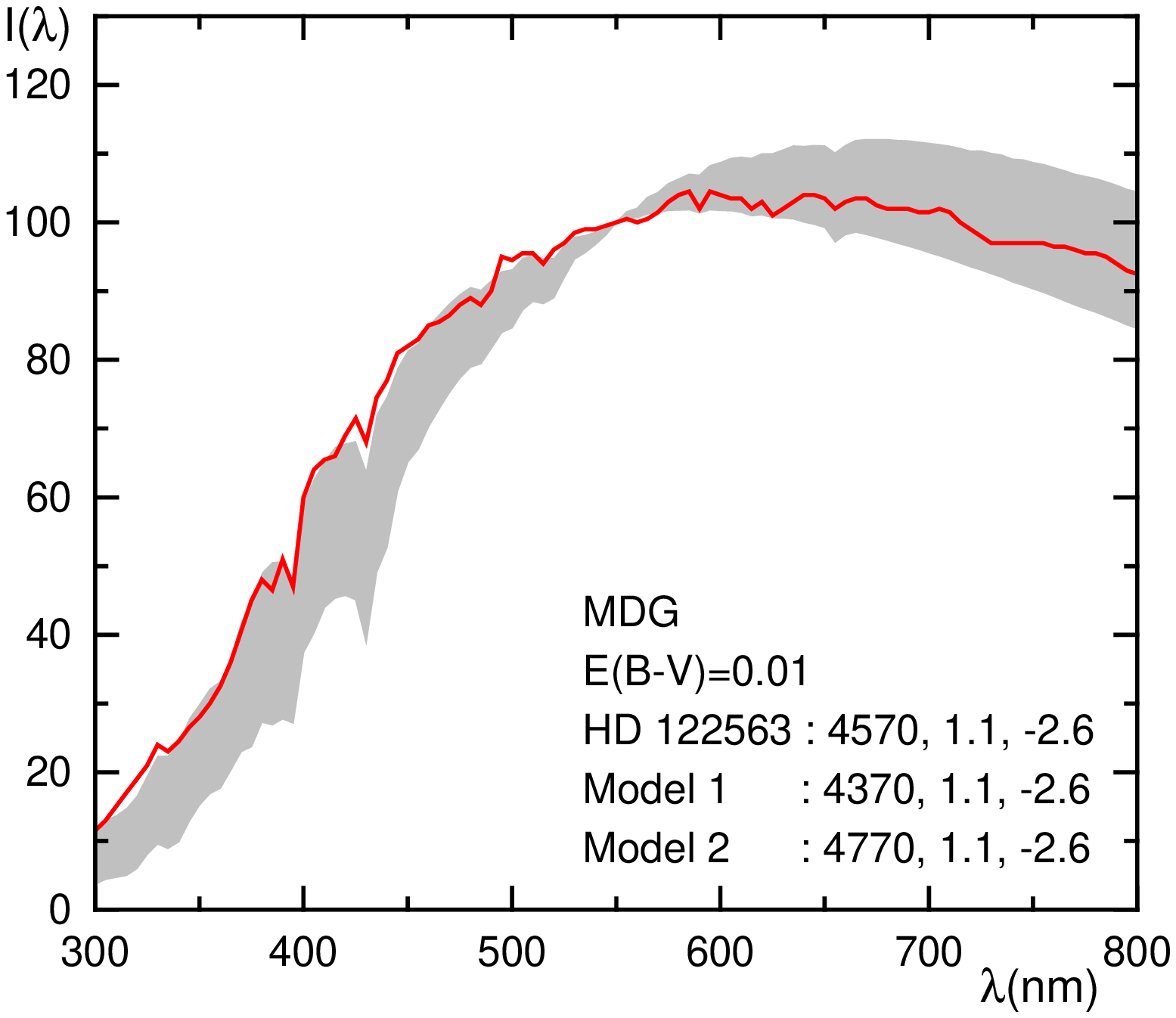,width=9.0truecm,angle=0,clip=}}
{ Spectral energy distribution of HD 122563 (MD giant). The effective
temperature of the model is changed by $\pm$200 K.}

\WFigure{22}{\psfig{figure=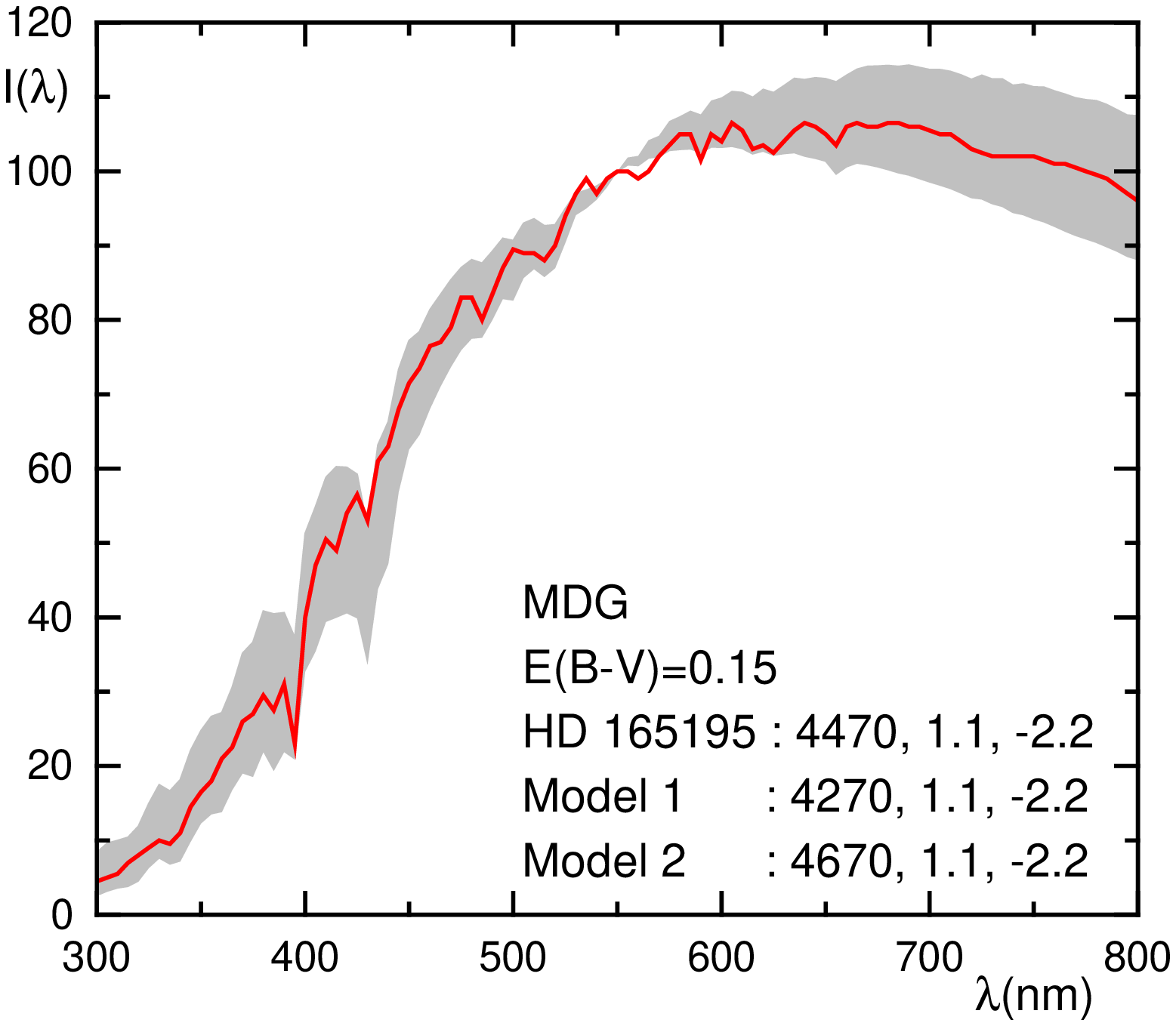,width=9.0truecm,angle=0,clip=}}
{ Spectral energy distribution of HD 165195 (MD giant). The effective
temperature of the model is changed by $\pm$200 K.}

\WFigure{23}{\psfig{figure=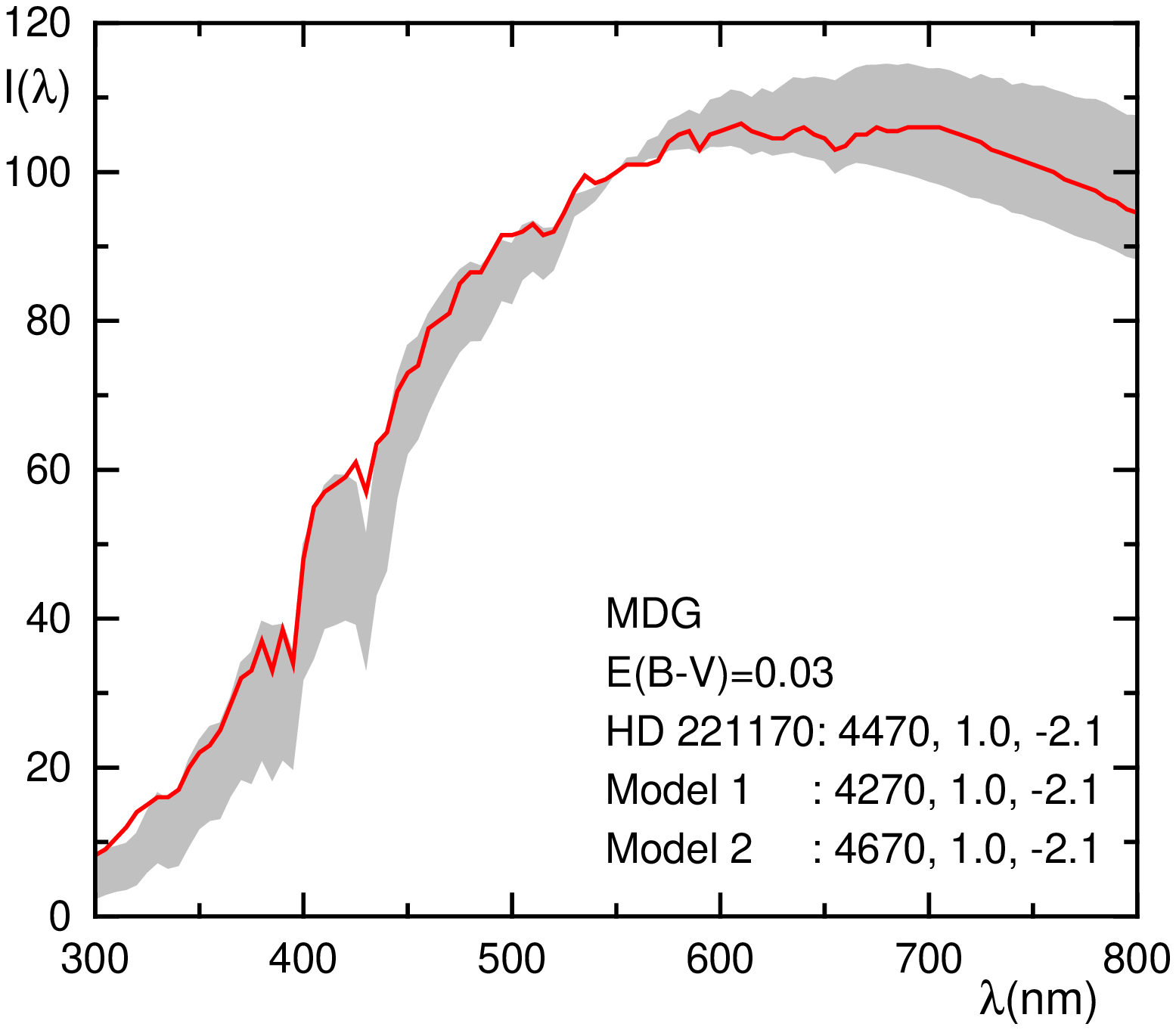,width=9.0truecm,angle=0,clip=}}
{ Spectral energy distribution of HD 221170 (MD giant). The effective
temperature of the model is changed by $\pm$200 K.}

\WFigure{24}{\psfig{figure=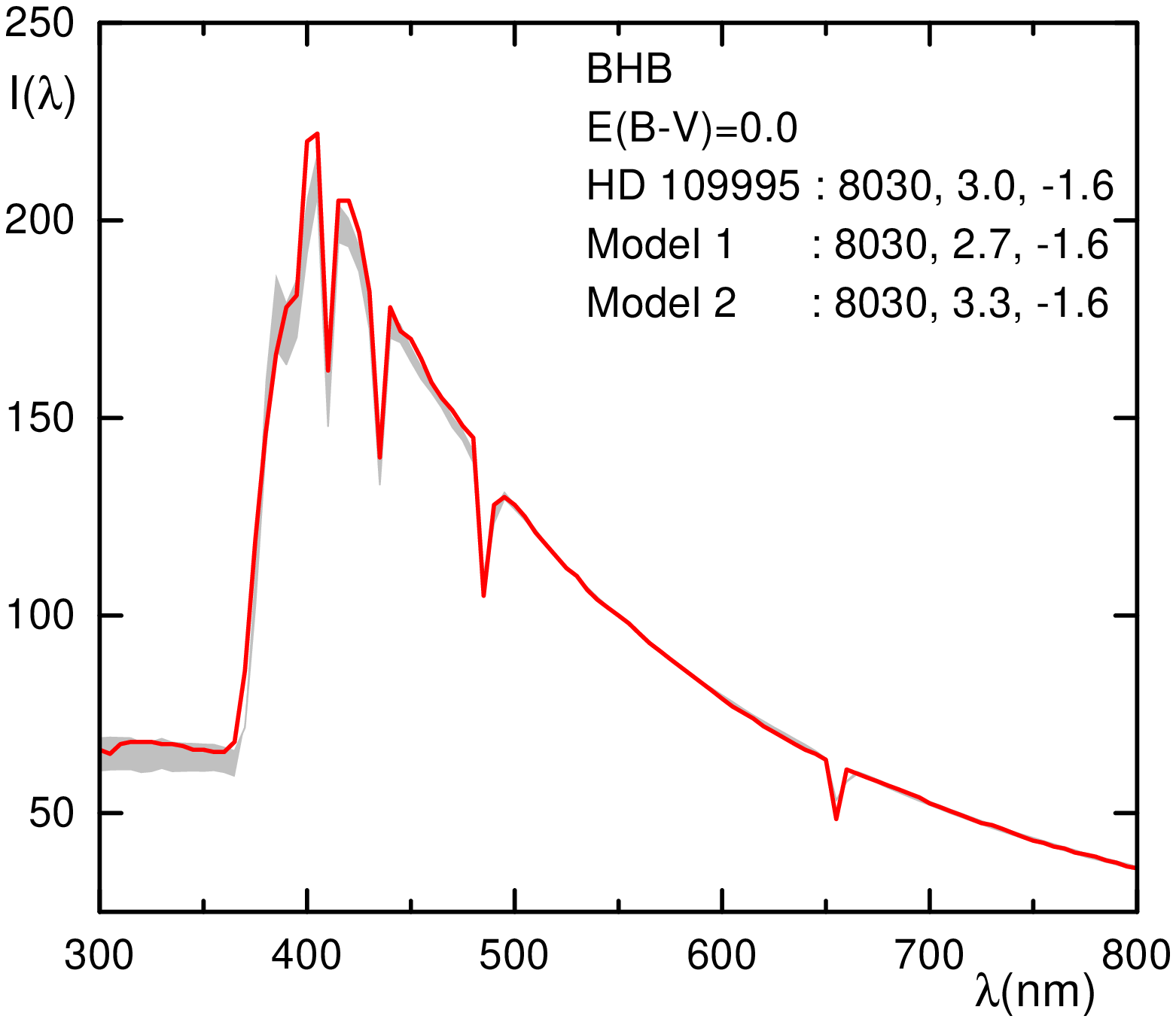,width=9.0truecm,angle=0,clip=}}
{ Spectral energy distribution of HD 109995 (BHB star). The gravity
of the model is changed by $\pm$0.3 dex.}

\WFigure{25}{\psfig{figure=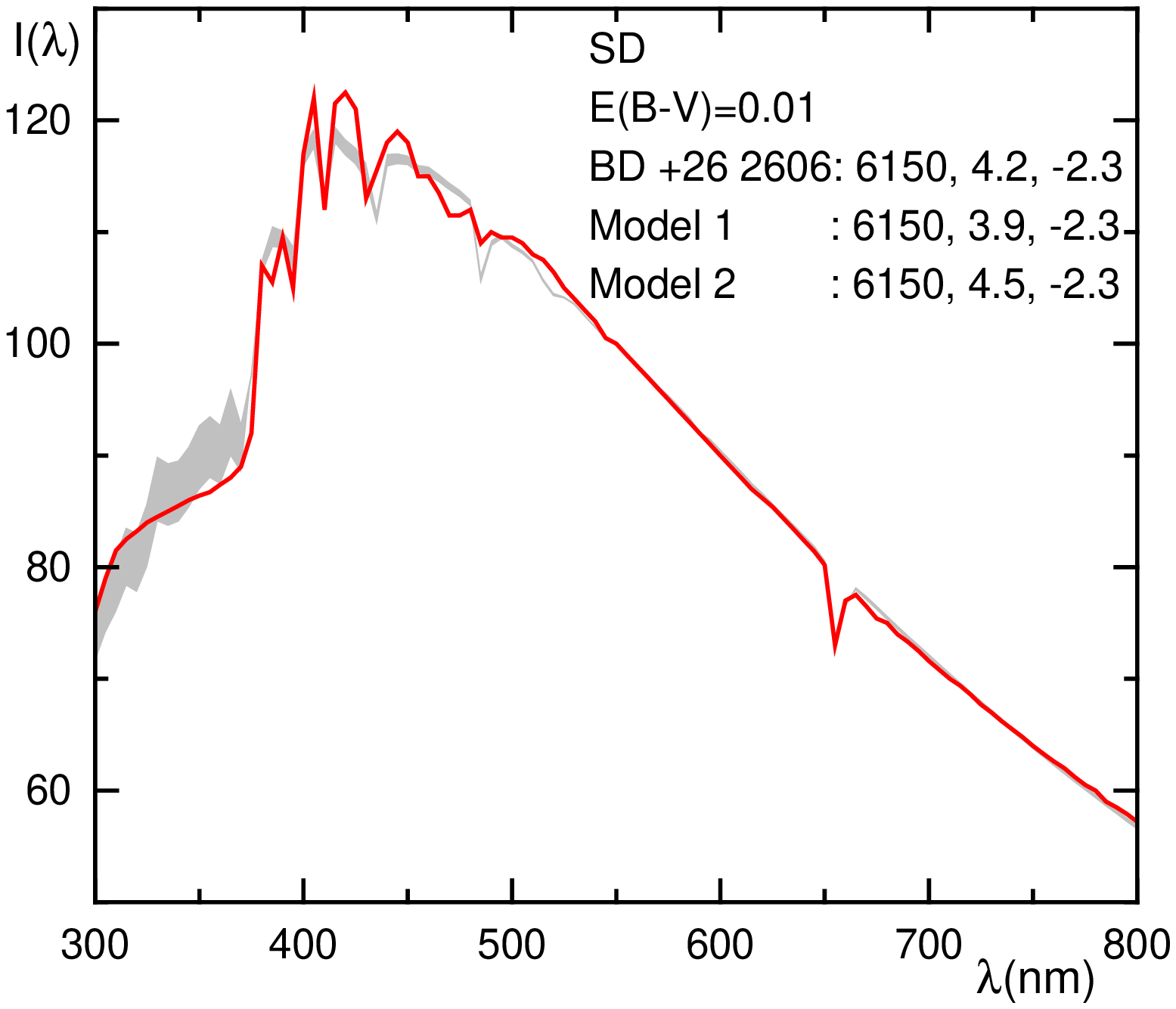,width=9.0truecm,angle=0,clip=}}
{ Spectral energy distribution of BD+26 2606 (subdwarf). The gravity
of the model is changed by $\pm$0.3 dex.}

\WFigure{26}{\psfig{figure=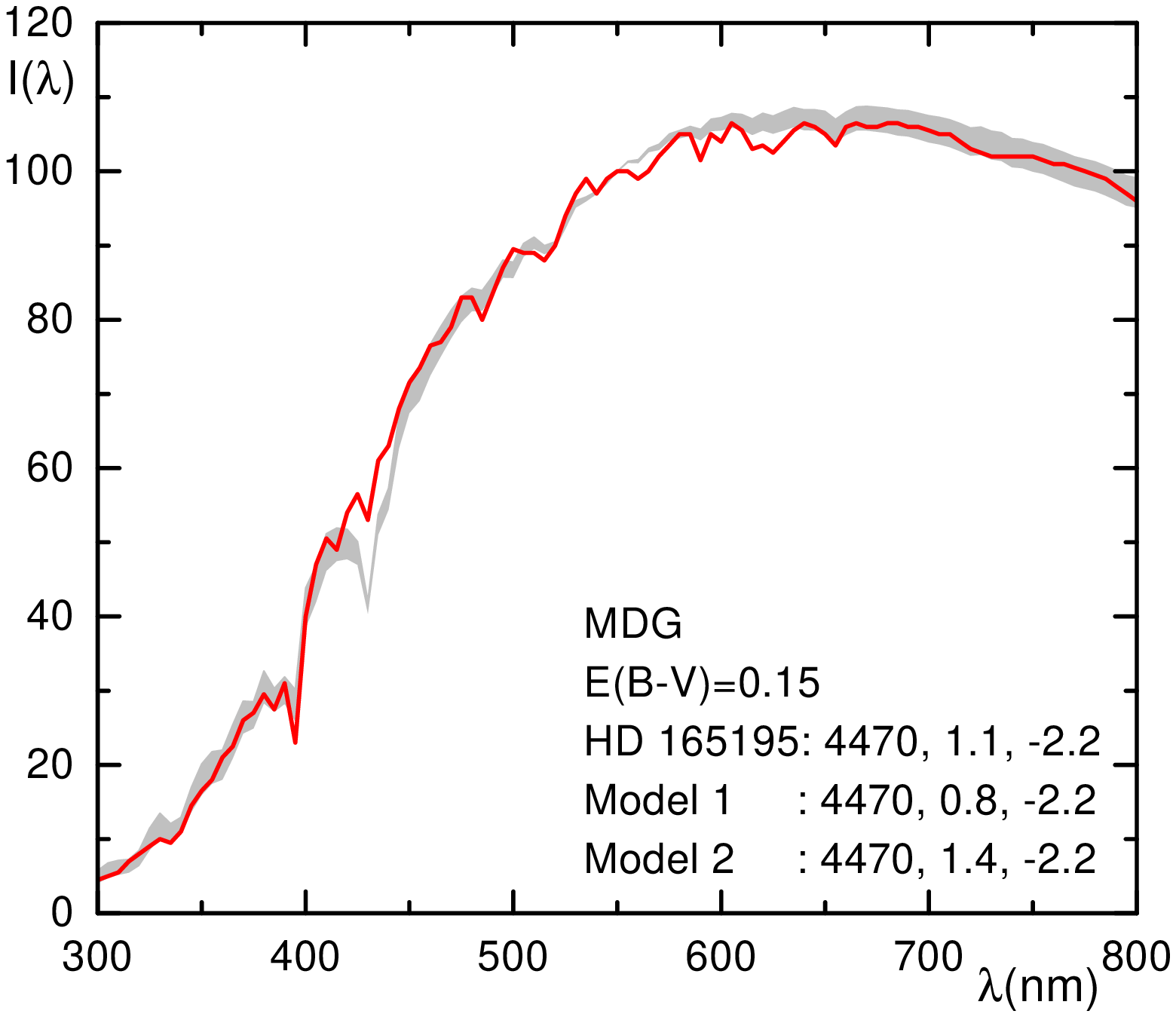,width=9.0truecm,angle=0,clip=}}
{ Spectral energy distribution of HD 165195 (MD giant). The gravity
of the model is changed by $\pm$0.3 dex.}

\WFigure{27}{\psfig{figure=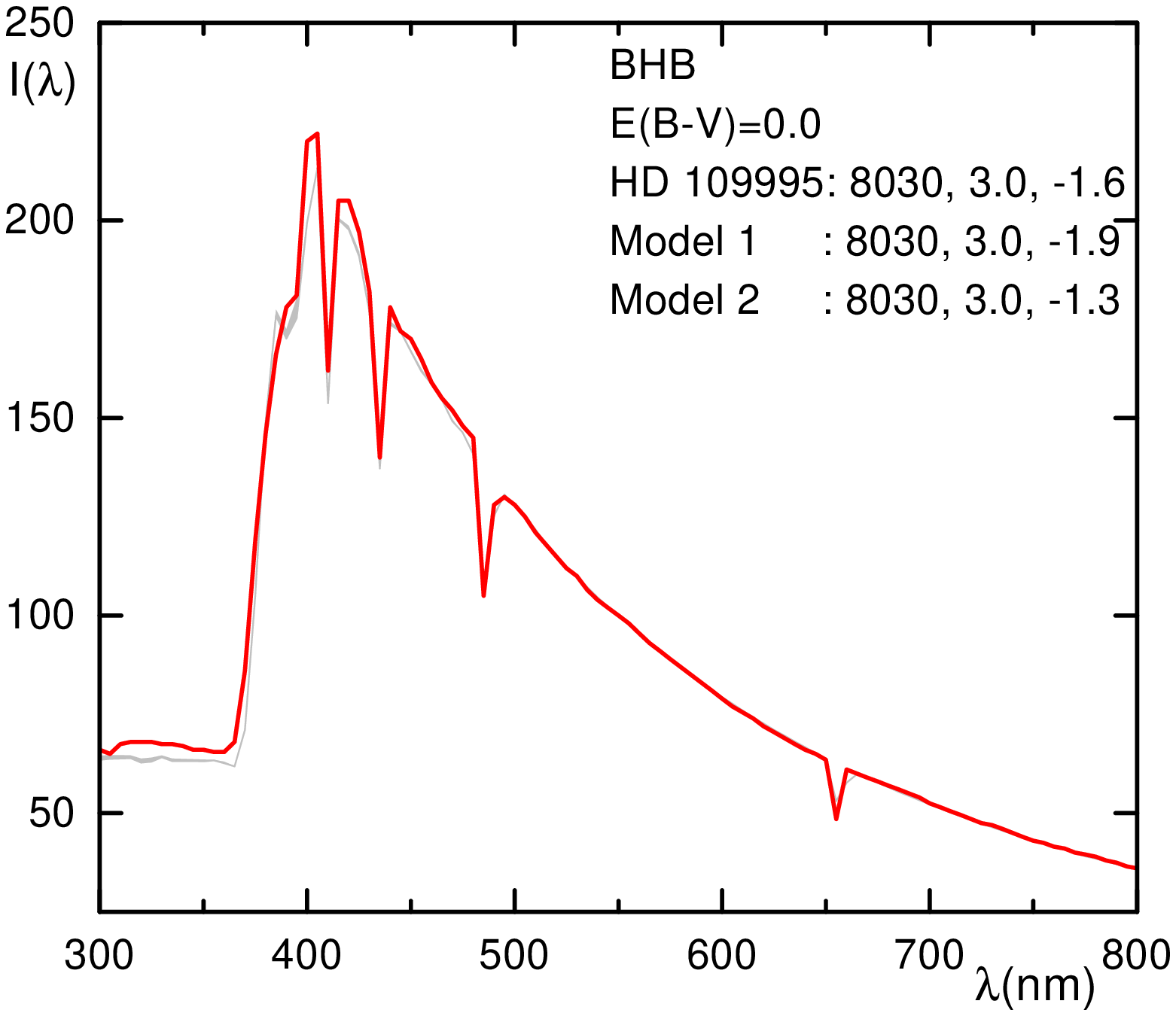,width=9.0truecm,angle=0,clip=}}
{ Spectral energy distribution of HD 109995 (BHB star). The metallicity
of the model is changed by $\pm$0.3 dex.}

\WFigure{28}{\psfig{figure=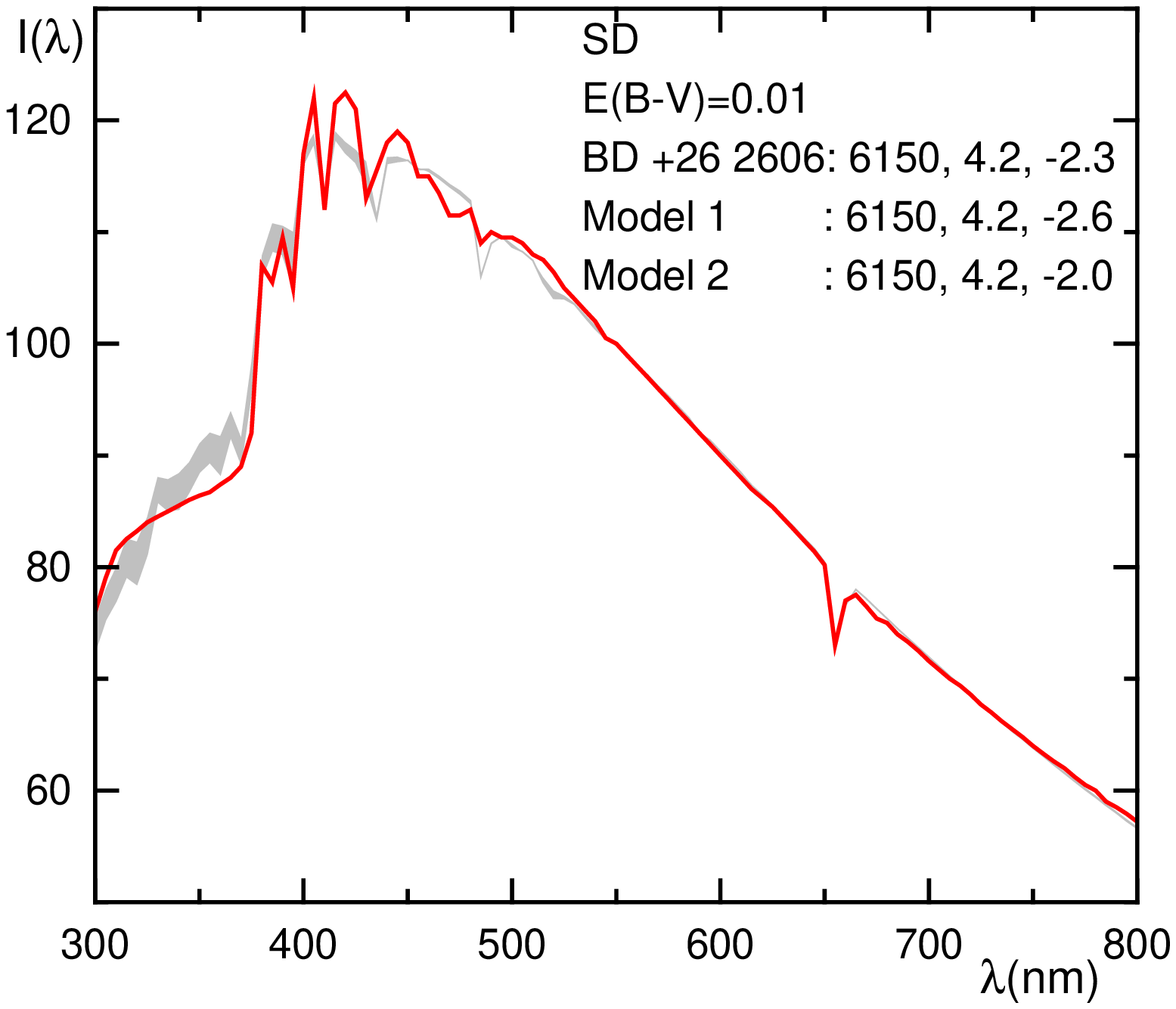,width=9.0truecm,angle=0,clip=}}
{ Spectral energy distribution of BD+26 2606 (subdwarf). The metallicity
of the model is changed by $\pm$0.3 dex.}

\WFigure{29}{\psfig{figure=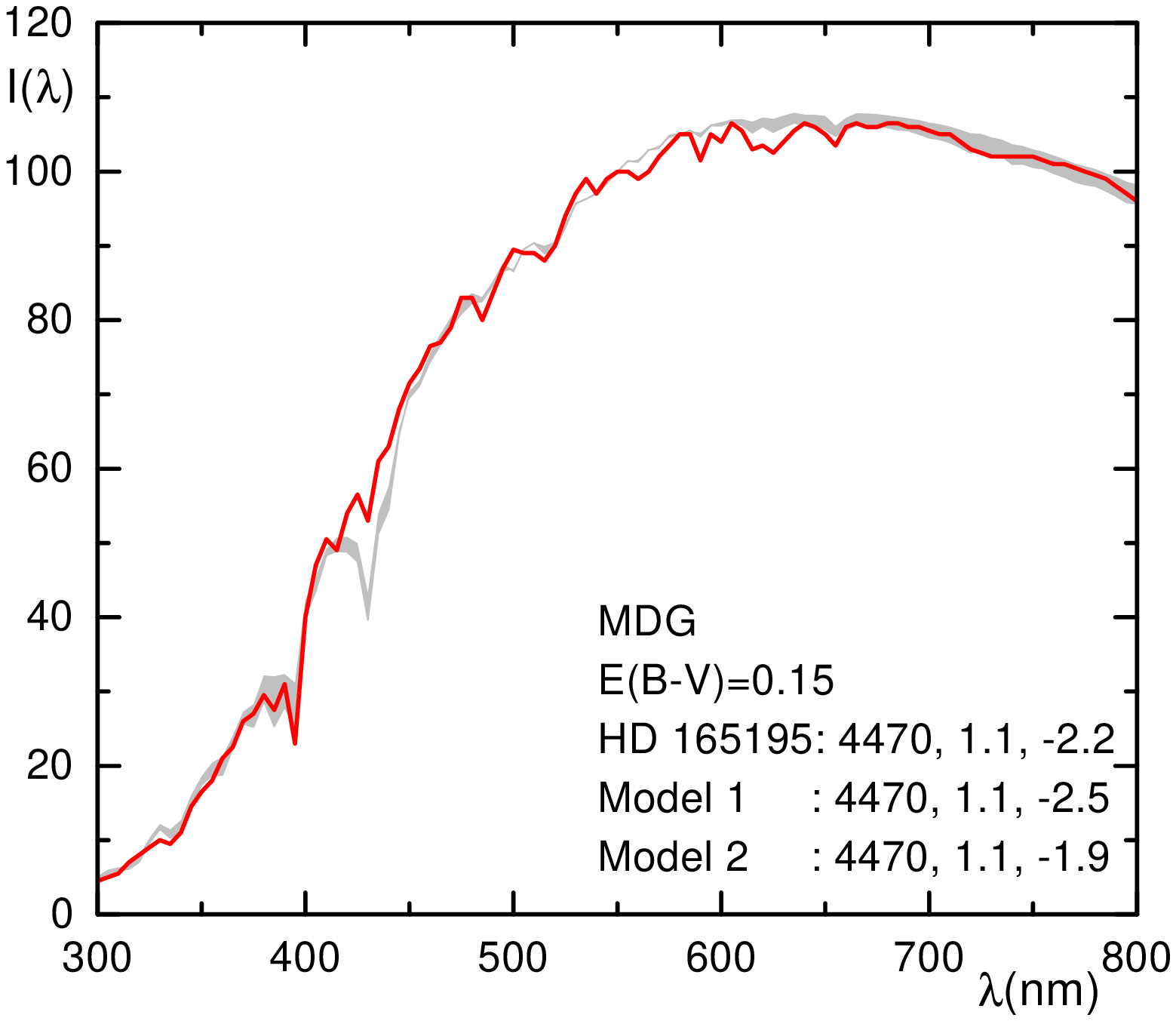,width=9.0truecm,angle=0,clip=}}
{ Spectral energy distribution of HD 165195 (MD giant). The metallicity
of the model is changed by $\pm$0.3 dex.}

\WFigure{30}{\psfig{figure=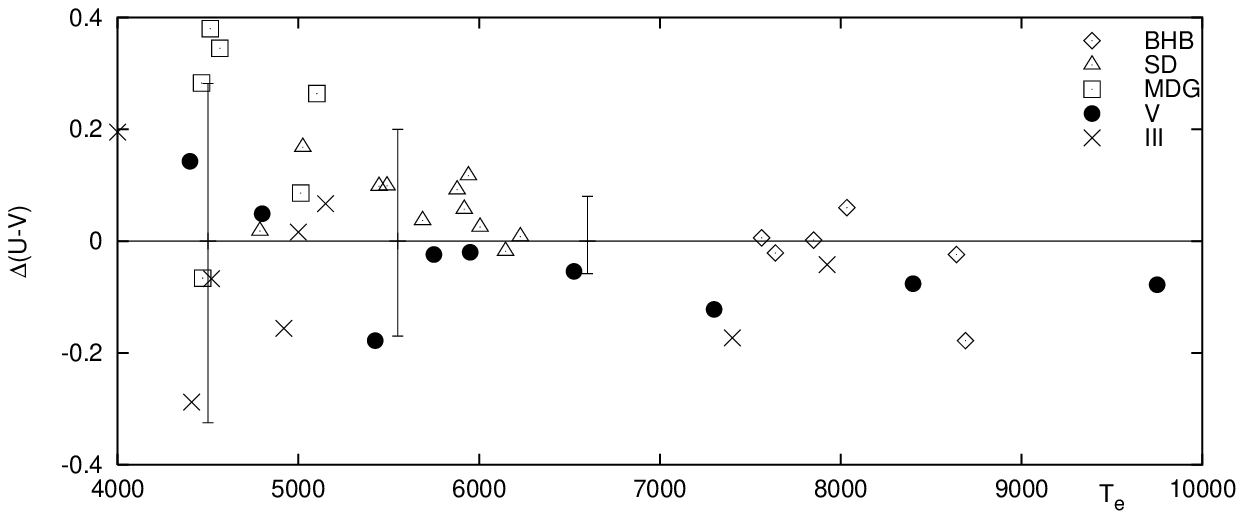,width=11truecm,angle=0,clip=}}
{ A comparison of the {\itl U}--{\itl V} color indices of stars and
their models (model minus observed). Mean wavelengths of the {\itl
U} and {\itl V} passbands are 345 and 544 nm. The vertical bars indicate
the errors originating from the temperature changes by $\pm$200 K.}

\WFigure{31}{\psfig{figure=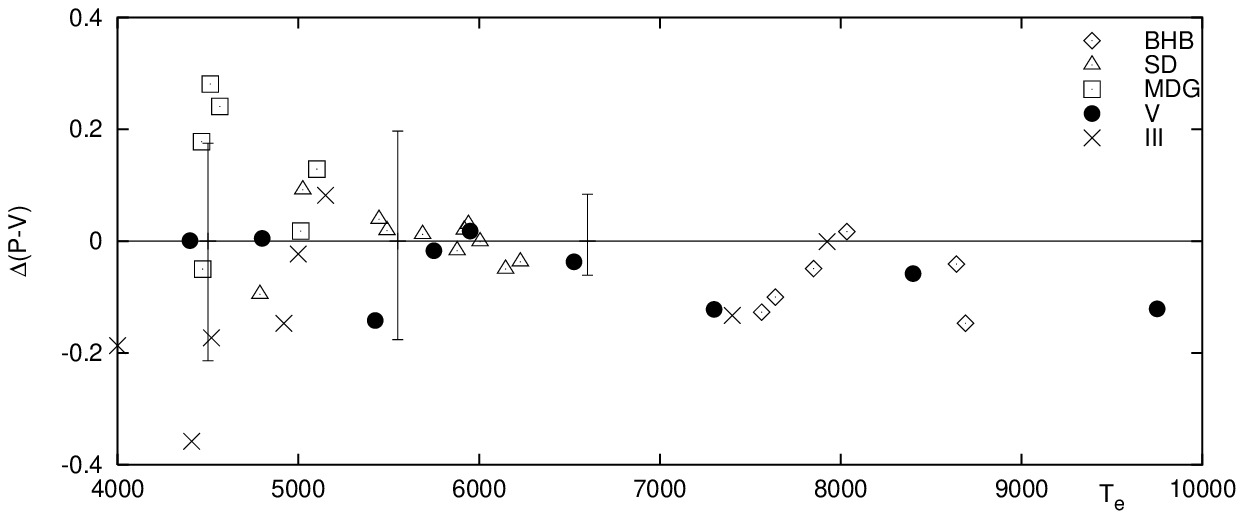,width=11truecm,angle=0,clip=}}
{ The same as in Fig. 30 but for {\itl P}--{\itl V}.
The mean wavelength of the {\itl P} passband is 375 nm.}

\WFigure{32}{\psfig{figure=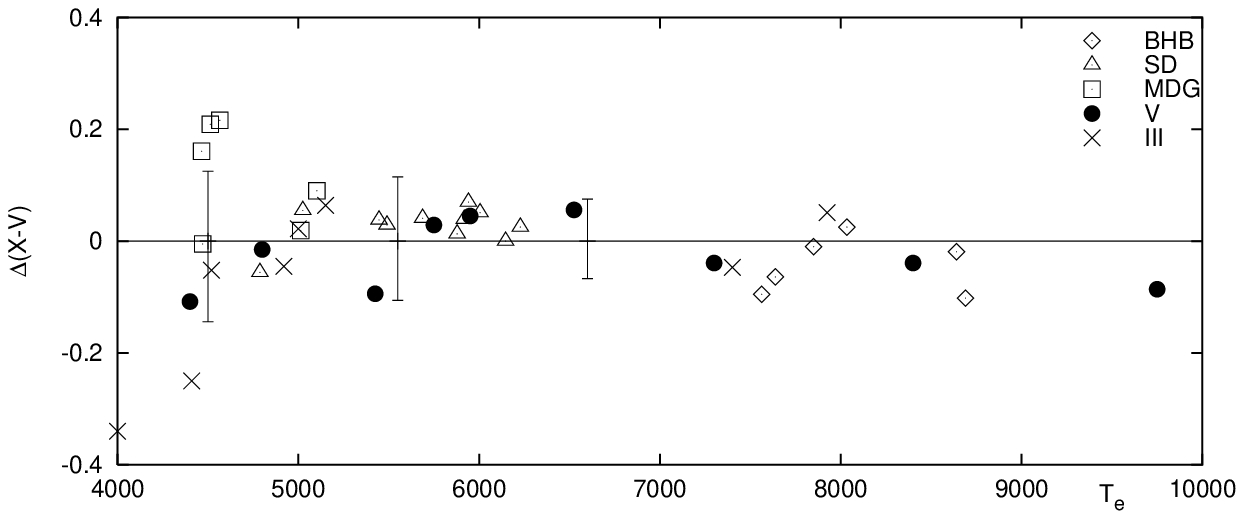,width=11truecm,angle=0,clip=}}
{ The same as in Fig. 30 but for {\itl X}--{\itl V}.
The mean wavelength of the {\itl X} passband is 405 nm.}

\WFigure{33}{\psfig{figure=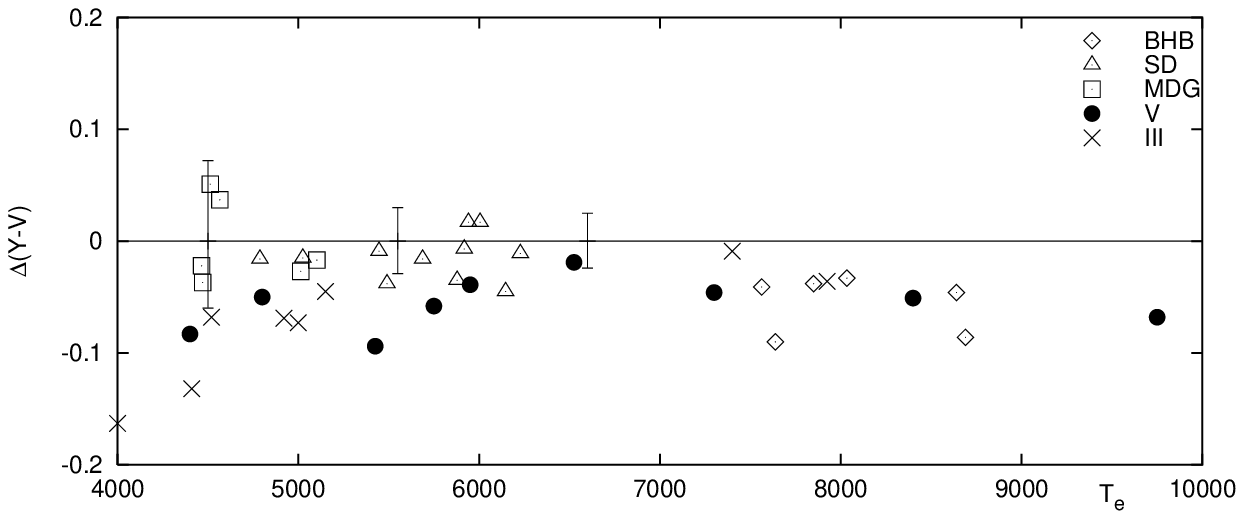,width=11truecm,angle=0,clip=}}
{ The same as in Fig. 30 but for {\itl Y}--{\itl V}.
The mean wavelength of the {\itl Y} passband is 466 nm.}

\WFigure{34}{\psfig{figure=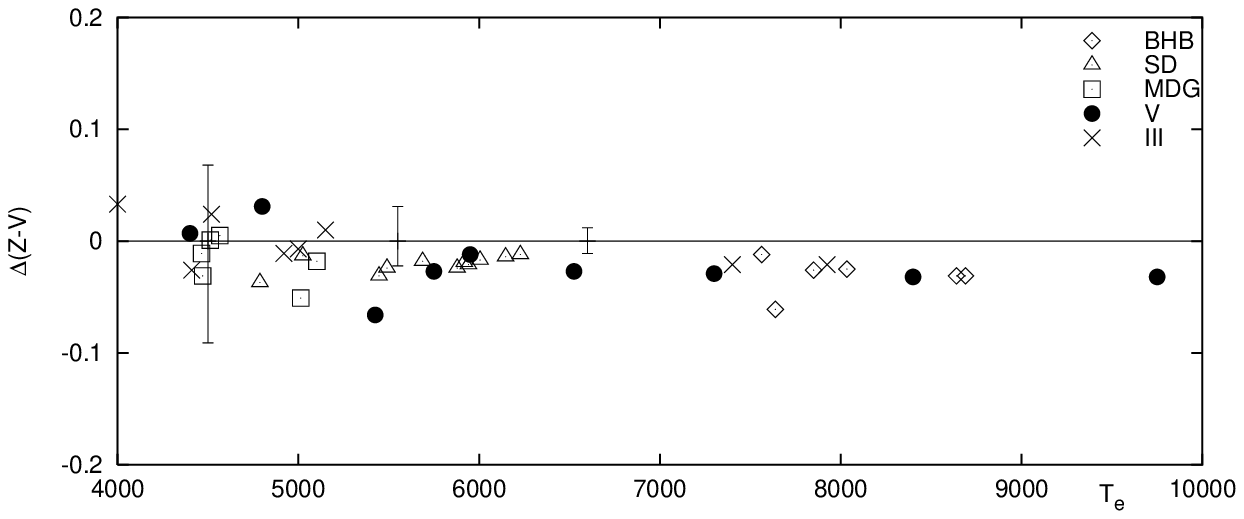,width=11truecm,angle=0,clip=}}
{ The same as in Fig. 30 but for {\itl Z}--{\itl V}.
The mean wavelength of the {\itl Z} passband is 516 nm.}

\WFigure{35}{\psfig{figure=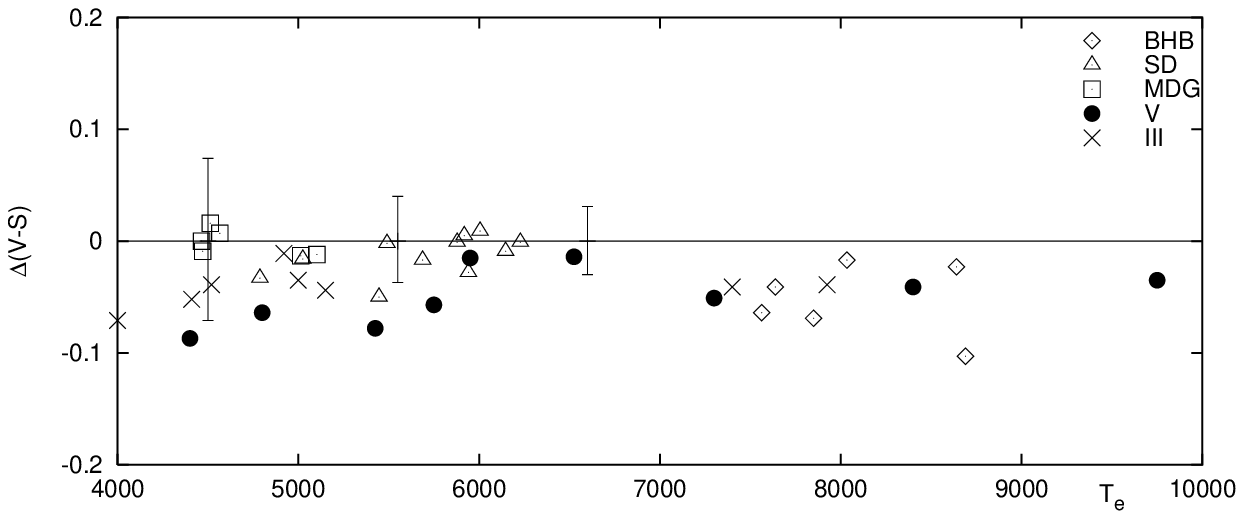,width=11truecm,angle=0,clip=}}
{ The same as in Fig. 30 but for {\itl V}--{\itl S}.
The mean wavelength of the {\itl S} passband is 656 nm.}

\WFigure{36}{\psfig{figure=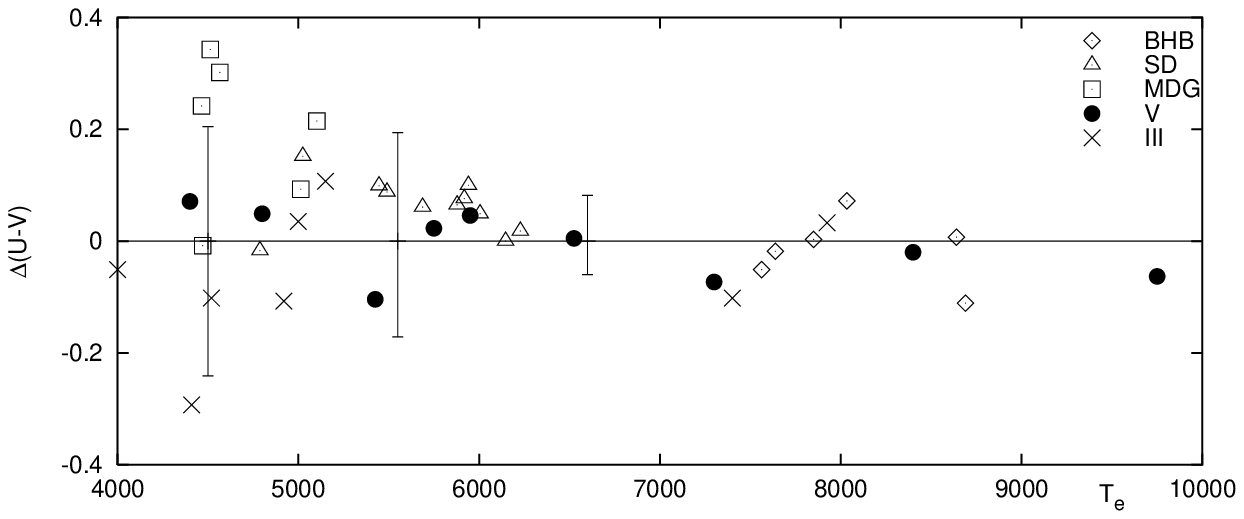,width=11truecm,angle=0,clip=}} {
The same as in Fig. 30 but for {\itl U}--{\itl V} of the {\itl UBV}
system.  Mean wavelengths of the {\itl U} and {\itl V} passbands are 364
and 550 nm.}

\WFigure{37}{\psfig{figure=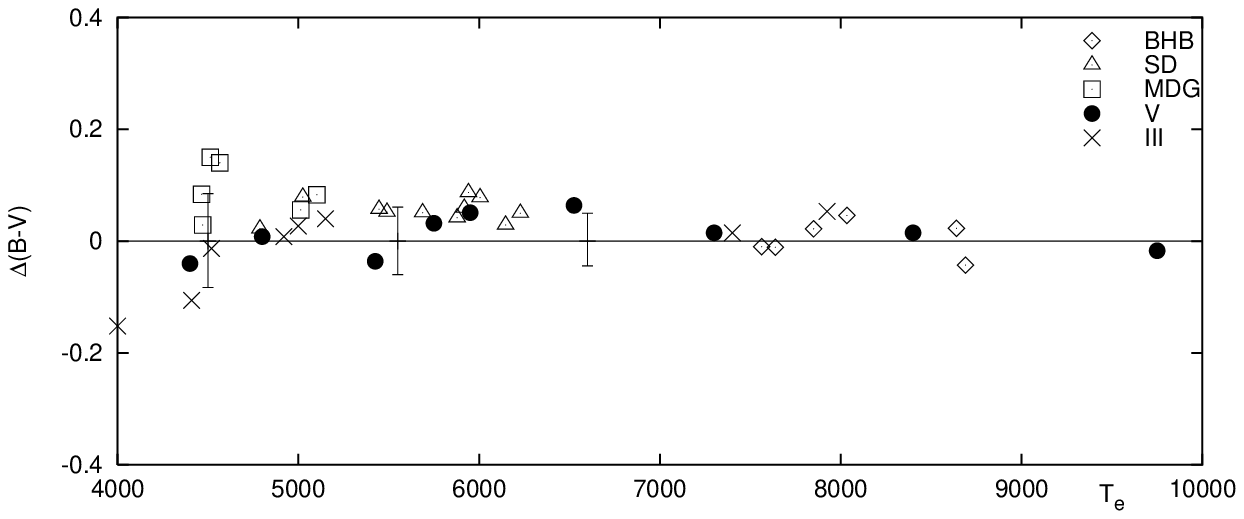,width=11truecm,angle=0,clip=}} {
The same as in Fig. 30 but for {\itl B}--{\itl V} of the {\itl UBV}
system.  The mean wavelength of the {\itl B} passband is 442 nm.}

\bye